\documentclass[twocolumn,showpacs,preprintnumbers,prb]{revtex4-1}
\usepackage{dcolumn}
\usepackage{amsmath}
\usepackage{amsfonts}
\usepackage{amssymb}
\usepackage{calligra}
\usepackage{bm}
\usepackage{graphicx}
\usepackage{float}
\usepackage{subfigure}
\usepackage{stackengine}

\newcommand{\e}[1]{e^{#1}} % for average
\newcommand{\avg}[1]{\left< #1 \right>} % for average
\newcommand{\abs}[1]{| #1 |} % for absolute value
\newcommand{\ket}[1]{| #1 \rangle} % for Dirac bras
\newcommand{\bra}[1]{\langle #1 |} % for Dirac kets

\DeclareMathAlphabet{\mathcalligra}{T1}{calligra}{m}{n}

\begin{document}

\title{Multi-fractality of the entanglement Hamiltonian eigen-modes}

\author{Mohammad Pouranvari}
\affiliation {Department of Physics, Faculty of Basic Sciences,
  University of Mazandaran, P. O. Box 47416-95447, Babolsar, Iran}

% \pacs{}
\date{\today}

\begin{abstract}
  We study the fractal properties of single-particle eigen-modes of
  entanglement Hamiltonian in free fermion models. One of these modes
  that has the highest entanglement information and thus called
  maximally entangled mode (MEM) is specially considered. In free
  Fermion models with Anderson localization, fractality of MEM is
  obtained numerically and compared with the fractality of Hamiltonian
  eigen-mode at Fermi level. We show that both eigen-modes have
  similar fractal properties: both have same single fractal dimension
  in delocalized phase which equals the dimension of the system, and
  both show multi-fractality at phase transition point. Therefore, we
  conclude that, fractal behavior of MEM -- in addition to the fractal
  behavior of Hamiltonian eigen-mode -- can be used as a quantum phase
  transition characterization.
\end{abstract}

\maketitle

\section{Introduction}\label{Introduction}

Quantum phase transition happens at zero temperature, where quantum
fluctuations -- in contrast to temperature dependent fluctuations --
are dominant and drive the phase transition. These fluctuations yield
to broadly distributed observable quantities at the phase transition
point. Anderson transition\cite{ref:andersonLDPTP} between delocalized
and localized phases, is one example of quantum phase transition which
has attracted much attention. The original Anderson phase transition
was introduced as a three-dimensional ($3d$) tight binding lattice
model with randomness in on-site energies. For a specific value of the
randomness strength, the state at the Fermi level becomes
localized. In this theory the lower critical dimension is $3$. In $1d$
and $2d$ cases, all states in the thermodynamic limit are localized
for an infinitesimal amount of disorder and subsequently there is no
Anderson phase transition. Later on, $1d$ and $2d$ models were
proposed with \textit{correlated disorder} that have
delocalized-localized transition.\cite{ref:mirlinreview} Anderson
transition, as a quantum phase transition, exhibits statistical
fluctuations at the phase transition point. These fluctuations are
manifested in the anomalous scaling of the inverse participation ratio
(defined below) of Hamiltonian eigen-mode at the Fermi energy
$\ket{E_F}$, which lead to fractal behavior at the phase transition
point. Such fractal behavior can be used as a tool to distinguish
different phases.\cite{ref:wegner} Multi-fractal analysis has broad
applications in different branches of science including
physiology,\cite{ref:stanley, ref:ivanov} geophysics,\cite{ref:davis,
  ref:tessier} fluid dynamics,\cite{ref:biferale, ref:meneveau,
  ref:schertzer, ref:prasad, ref:meneveau2, ref:benzi} and even in
finance.\cite{ref:zunino, ref:matia, ref:sun, ref:oswie, ref:oswie2}

Some early reports on the fractal behavior of the $\ket{E_F}$ are
Refs. [\onlinecite{ref:zdetsis, ref:soukoulisFractal, ref:castellani,
  ref:aoki, ref:falko, ref:mildenberger, ref:janseen}] where its
fractal properties are analyzed in different phases and used as a
phase characterization. Some more recent reports are the followings:
based on an analytic calculation, Ref. [\onlinecite{ref:mirlin}] shows
that there is a symmetry in the multi-fractal spectrum of
$\ket{E_F}$. The relation between single-particle entanglement entropy
and fractal dimensions at the phase transition point was found in
Refs. [\onlinecite{ref:jia, ref:giraud}]; also
Ref. [\onlinecite{ref:chen}] proves analytically and then shows
numerically that multi-fractality of $\ket{E_F}$ at the phase
transition point can be obtained by using moments of R\'{e}yni
entropy. Refs. [\onlinecite{ref:vasquez, ref:rodriguez2008}] calculate
the singularity spectrum of $\ket{E_F}$ in Anderson $3d$ model and
compare the typical average with the ensemble average in calculation
of singularity spectrum. Furthermore, in Ref.
[\onlinecite{ref:rodriguez2009}] singularity spectrum is obtained by
calculating the probability distribution of $\abs{\psi_i}^2$ ($\psi_i$
as the wave-function at site $i$) .

In this paper, we propose to study the problem of Anderson
transition from the point of view of multi-fractality of
entanglement Hamiltonian. Let us recall the definition of
entanglement Hamiltonian. If the system is in state $\ket{\psi}$,
then the density matrix will be given by $\rho=\ket{\psi}
\bra{\psi}$. For a bipartite system, reduced density matrix for
one subsystem is obtained by tracing over degrees of freedom of
other subsystem. As we know, for free Fermion, we can write the
reduced density matrix as $\e{-H_{ent}}$, where the $H_{ent}$ is a
free Fermion Hamiltonian and called entanglement Hamiltonian.
Entanglement Hamiltonian eigen-modes of two subsystems are then
attached together to make a mode for the entire
system.\cite{ref:klich} The mode corresponding to smallest
magnitude entanglement energy which has the largest contribution
to the entanglement entropy is distinguished from others, since it
has important physical information about the
system.\cite{ref:pouranvariyang2, ref:pouranvariyang1,
  ref:pouranvariyang5, ref:pouranvariafshin2016} This mode is called maximally
entangled mode (MEM).

In Ref. [\onlinecite{ref:pouranvariyang2}] it is shown, regarding the
localization of the mode, that MEM and $\ket{E_F}$ contain the same
physics: both are localized in the localized regime and both are
extended in delocalized regime.  In addition, their overlap at the
phase transition point is larger than their overlap in delocalized or
localized phases, although small compare to $1$. Here, regarding the
comparison of two modes from indirect point of view of fractality, we
ask the following questions: does MEM show multi-fractality at the
phase transition point? Can we use fractal behavior of MEM to
distinguish different phases?

In this paper, we answer the above questions, using two $1d$ models
and the $3d$ Anderson model. We obtain $\ket{E_F}$ using numerically
exact diagonalization of the Hamiltonian which is an $N\times N$
matrix ($N$ is the system size). And to obtain the MEM, we follow the
method mentioned in Ref. [\onlinecite{ref:klich}] where we have to
diagonalize another matrix with dimension $N_F\times N_F$ ($N_F$ is
the Fermion number). These two diagonalization procedures make the
calculations very time consuming and thus we are limited in the system
size for the case of $3d$ Anderson model.

Our key results are as follows: In the delocalized phase, MEM, like
$\ket{E_F}$ has a single fractal dimension equal to dimension of the
system $d$, while in the localized phase, the fractal dimension goes
to zero. More importantly, at the phase transition point, MEM shows
multi-fractality; we calculated numerically its multi-fractal spectrum
and also show that MEM obeys the symmetry relation of anomalous
exponents. Furthermore, we can distinguish different phases based on
the singular spectrum of the MEM.

This paper is organized as follows: in section \ref{MF} we explain
multi-fractality as a mathematical concept and then apply it to
wave-function in lattice systems. The models we intend to study are
next explained in section \ref{models}. Section \ref{MMEM} contains
main results of our numerical calculations. Finally, the summary of
our work is presented in section \ref{conc}.

\section{Multi-fractality analysis}\label{MF}
Suppose that we have $\mathcal{N}$ numbers, randomly
distributed. Dividing this set of numbers into cells with size $\ell$,
the probability that a number is in the $i$th cell, $p_i(\ell)$ is
proportional to the numbers included in that cell $\mathcal{N}_i$:
$ p_i(\ell) = \mathcal{N}_i / \mathcal{N}.$

Scaling behavior of moments of the probability, averaged over all
cells, tells us the multi-fractal structure of these random numbers:
\begin{equation}
  \avg{p_i(\ell)^{q-1}} \propto \ell^{\tau(q)},
\end{equation}
where the multi-fractal spectrum is defined as below:
\begin{equation}
  \tau(q) = (q-1)D(q).
\end{equation}
If $D(q)$ is independent of $q$, we call $D$ the single-fractal
dimension; otherwise, when $\tau$ is not a linear function of $q$, we
have \emph{multi}-fractality.\cite{ref:castellani}

Now, in view of above method of characterizing random numbers, the
fractal behavior of an eigen-function in a lattice system can be
studied, where $|\psi_i|^2$'s for the $i$th lattice sites are the
random numbers. We want to obtain the scaling behavior of the so
called generalized inverse participation ratio(GIPR) $P_q$, defined below:
\begin{eqnarray}\label{Ppsi}
  P_q(\ell) &=& \sum_k^{N_{\ell}} \mu_k^q(\ell), \\
  \mu_k(\ell) &=& \sum_i^{\ell} |\psi_i|^2,\label{Ppsi2}
\end{eqnarray}
in which we divide the system with size $N$ into $N_{\ell}$ cells,
each containing $\ell$ sites and we coarse grain over cells with Eq. (\ref{Ppsi2}). For a
wave-function
\begin{equation}\label{pq}
P_q \sim \lambda^{\tau(q)}, \quad \text{where} \ \lambda =
\ell/N.
\end{equation}

The behavior of the multi-fractal spectrum $\tau(q)$ can be used as a
characterization for Anderson localization:\cite{ref:wegner}
\begin{equation}\label{tau}
  \tau(q) \sim
\begin{cases}
    0          ,& \text{in localized phase} \\
    D(q)(q-1), & \text{at the phase transition point} \\
    d(q-1),    &\text{in delocalized phase}
\end{cases}
\end{equation}
i.e. in the localized phase no scaling behavior is seen. In the
delocalized phase, the singularity spectrum $\tau$ is a linear
function of $q$ with a constant slope of $d$ and thus the
wave-function is considered to have \textit{single-fractal}
dimension. On the other hand, at the phase transition point, $\tau(q)$
is a non-linear function of $q$ with a varying slope of $D(q)$ and the
wave-function is \textit{multi-fractal}.

In addition, $\tau(q)$ is written as:
\begin{equation}\label{Delta}
  \tau(q) = d(q-1) + \Delta_q,
\end{equation}
where $\Delta_q$ are the anomalous exponents that are zero in the
delocalized phase and hold the following symmetry relation at the
phase transition point:\cite{ref:mirlin}
\begin{equation}\label{symDelta}
  \Delta_q = \Delta_{1-q}.
\end{equation}

By applying Legendre transformation, one obtains the singularity
spectrum $f(\alpha)$:
\begin{eqnarray}\label{falpha}
  \alpha &=& \frac{d\tau(q)}{dq}, \\
  f(\alpha) &=& q \frac{d \tau}{dq} - \tau. \label{falpha2}
\end{eqnarray}

$f(\alpha)$ is the \emph{fractal} dimension of points where
$|\psi_i|^2 = N^{-\alpha}$, i.e. number of such points that scale as
$N^{f(\alpha)}$.

\section{models}\label{models}
The first model we study is the Aubry-Andre (AA)
model.\cite{ref:aubryandre} It is a $1d$ tight binding model with the
Hamiltonian:
\begin{equation}\label{oh}
  H=-t\sum_{<i,j>} (c_i^{\dagger} c_{j}+c_{j}^{\dagger} c_i)+
  \sum_{i}\epsilon_i c_i^{\dagger} c_i,
\end{equation}
where $c^{\dagger}_i(c_i)$ is the creation (annihilation) operator for
the site $i$ in the second quantization representation and $<>$
indicates nearest neighbor hopping only. Hopping amplitudes are
constant $t=1$, and on-site energy $\epsilon_i$ at site $i$ has an
incommensurate period:
\begin{equation}\label {AAHamiltonian}
  \epsilon_i = 2 \eta \cos{(2\pi i b)},
\end{equation}
where $b = \frac{1+\sqrt{5}}{2}$ is the golden ratio. This model has a
phase transition at $\eta=1$. As we change $\eta$, we go through
a phase transition from delocalized states ($\eta < 1$) to
localized states ($\eta >1$).

Another model is power-law bond disordered Anderson model
(PRBA)\cite{ref:lima} which is a $1d$ model with the
Hamiltonian:
\begin{equation}\label{Hhij}
  H = \sum_{i,j=1}^N h_{ij}c^{\dagger}_i c_j
\end{equation}
in which on-site energies are zero, and long-range hopping amplitudes
are
\begin{equation}
  h_{ij}= w_{ij}/|i-j|^{a}
\end{equation}
where $w$'s are uniformly random numbers distributed between $-1$ and
$1$. There is a phase transition at $a=1$ between delocalized state
($a<1$) and localized state ($a>1$). The other model is power-law
random banded matrix model (PRBM)\cite{ref:mirlinprbm} which is a $1d$
long range hopping model with the Hamiltonian of Eq. (\ref{Hhij}):
matrix elements $h_{ij}$ are random numbers, distributed by a Gaussian
distribution function that has zero mean and the following variance
(with periodic boundary condition):
\begin{equation}
  \avg {\abs{h_{ij}}^2} = \left[{1+\left(\frac{\sin{\pi
(i-j)/N}}{b \pi /N}\right)^{2a}}\right]^{-1},
\end{equation}
The system is delocalized for $a<1$; at the phase transition point
$a=1$, it undergoes Anderson localization transition to localized
states for $a>1$. This phase transition happens regardless of $b$, and
in our calculation we set $b=1$. Specially this model is important for
us, since by changing parameter $b$, we can simulate different
models.\cite{ref:jose,ref:balatsky,ref:altshuler,ref:ponomarev,ref:casati}
Interestingly, it has similar multi-fractal properties like the
Anderson model in three dimensions.\cite{ref:mirlin2000}

And finally, we also use $3d$ Anderson model (Eq. (\ref{oh})) with
randomly Gaussian distributed on-site energies, $\epsilon_i$, and
constant nearest-neighbor hopping amplitudes, $t=1$. The Gaussian
distribution has zero mean and variance $w$. Anderson phase transition
happens at $w=6.1$, with delocalized behavior for $w<6.1$ and
localized behavior for $w>6.1$.\cite{ref:markos}

\section{Multi-Fractality of Maximally Entangled Mode}\label{MMEM}
Multi-fractal analysis of Hamiltonian eigen-mode at the Fermi energy
$\ket{E_F}$ has been studied before\cite{ref:aoki,
  ref:soukoulisFractal, ref:castellani, ref:zdetsis,
  ref:rodriguez2008, ref:vasquez}. Here, fractal
properties of MEM is studied and compared with the $\ket{E_F}$. To do
so, in the following we first inspect profile of MEM in AA model. Then
multi-fractal spectrum as well as the singularity spectrum of MEM in
PRBA, PRBM, and Anderson $3d$ are studied. Then, the symmetry relation
of the anomalous exponents $\Delta$, Eq. (\ref{symDelta}) for the MEM
is verified.

\subsection{Profile of MEM}
\begin{figure}
  \centering
  \includegraphics[width=0.48\textwidth]{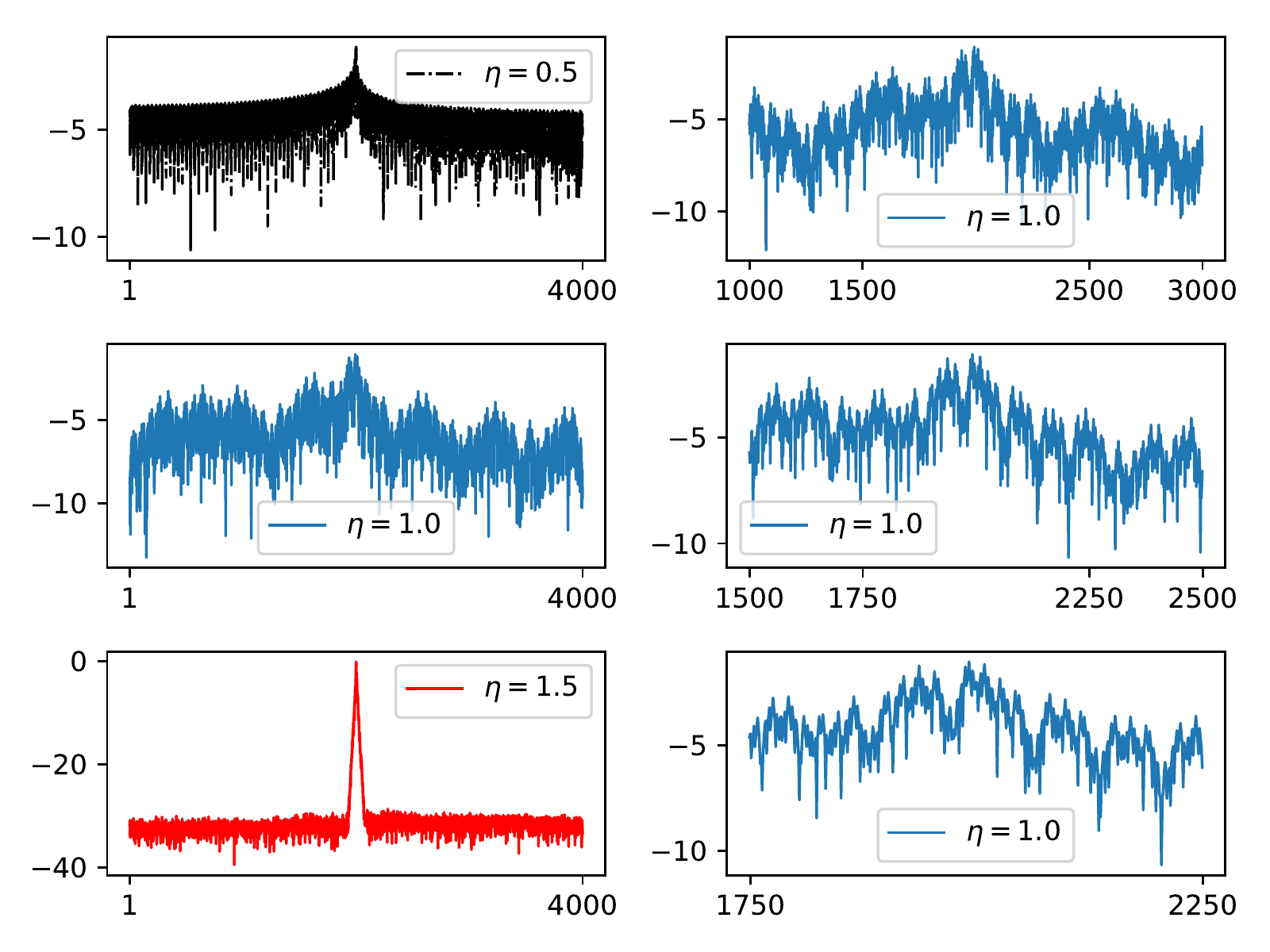}
  \caption{(color online) Left panels are $\log_{10} |\psi_i|^2$ of
    MEM for AA model versus site number for a sample with system size
    $N=4000$ in delocalized phase ($\eta=0.5$, top panel), at the
    phase transition point ($\eta=1.0$, middle panel), and in the
    localized phase ($\eta=1.5$, bottom panel). In each of the
    right panels, a different part of the MEM at the phase transition
    point is plotted. For each choice, we see the same behavior. Thus
    MEM at the phase transition point is self-similar. Note that there
    is no randomness in AA model and we do not have to take disorder
    average.
    \label{fig:MEM_AA_psi}}
\end{figure}

First, we look at the profile of MEM in different phases for AA model,
which is a disorder-free model and we do not have to take disorder
average. We plot MEM in the delocalized phase, localized phase and at
the phase transition point in Fig. \ref{fig:MEM_AA_psi}. As we can
see, in the delocalized phase, MEM is spread over sites, while it is
localized at one site in the localized phase. On the other hand, it
shows self-similarity at the phase transition point. i.e. behavior of
any part of the MEM is similar to that of the entire mode. Because of
such self-similarity, MEM shows multi-fractality.
%Self-similarity of $\ket{E_F}$ in AA model is shown in Fig. \ref{fig:UhNF_AA_psi}.

\begin{figure}
  \centering
  \includegraphics[trim = 0.2cm  5cm 0 0 clip,width=0.5\textwidth]{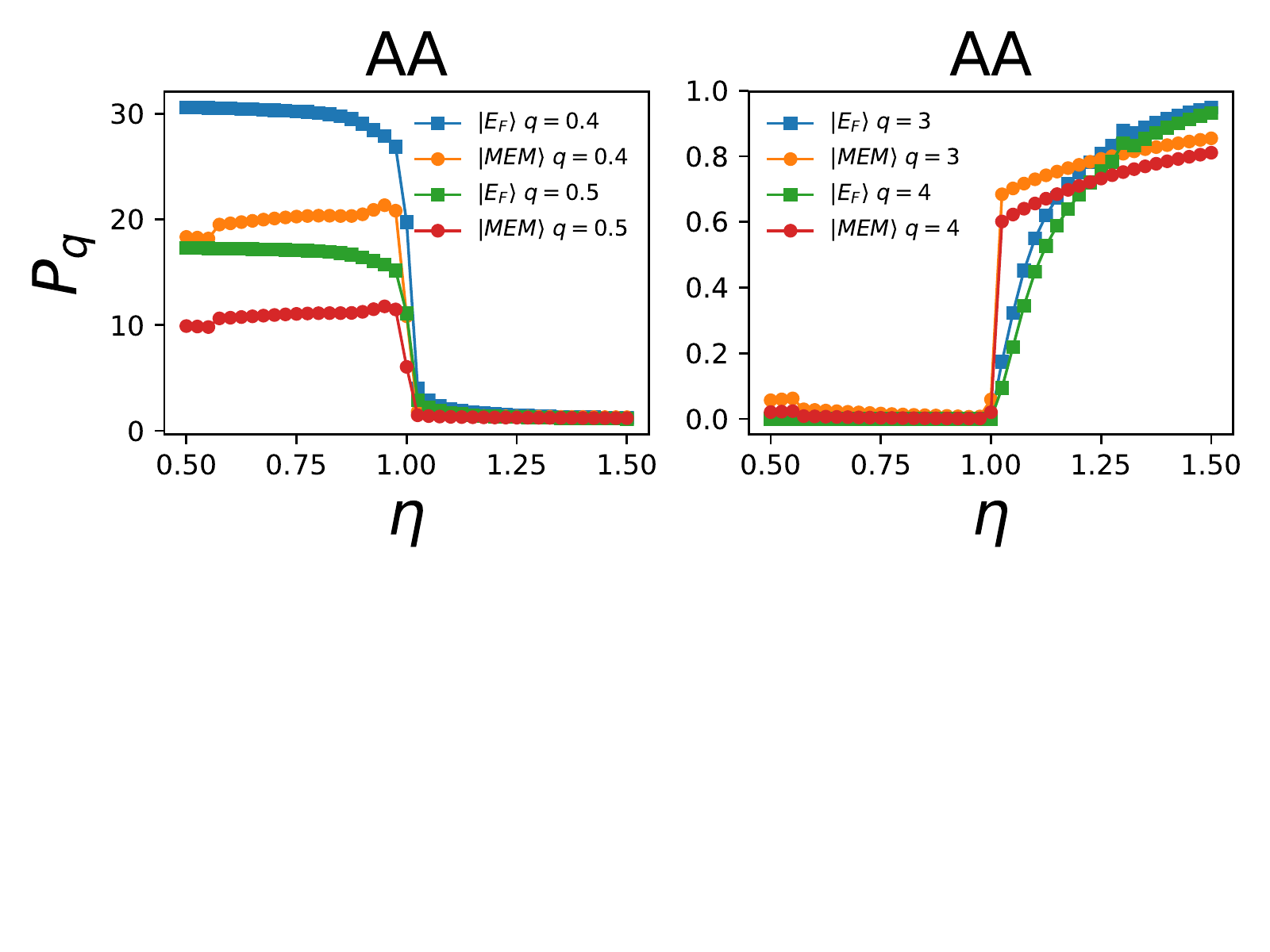}
  \caption{(color online) Generalized participation ratio,
    Eq. (\ref{Ppsi}) of both $\ket{E_F}$ and MEM, for the AA
    model. Left panel: $q=0.4, 0.5$ and Right panel: $q=3,4$. Both
    modes have similar behavior in delocalized phase ($\eta<1$) and in
    localized phase ($\eta>1$). System size is $N=3000$. Since there
    is no randomness in AA model, we do not have to take disorder
    average.  }
  \label{fig:P_AA_N_3000}
\end{figure}

Moreover, Since fractal properties of eigen-modes are extracted from
the GIPR, we compare the GIPR of MEM and $\ket{E_F}$ according to
Eq. (\ref{Ppsi}). GIPR of AA model for different $q$'s are plotted in
Fig. \ref{fig:P_AA_N_3000}. For each $q$, although the behavior is
not identical, similar trend is observed. From this simple
calculation, we can deduce that MEM has much the same fractal
properties as $\ket{E_F}$. In addition, similar to the GIPR of the
Hamiltonian eigen-mode at the Fermi level, GIPR of the MEM
distinguishes different phases and can be used as a phase detection
parameter.

\subsection{Multi-fractal Spectrum}
In this subsection, we consider the behavior of multi-fractal spectrum
$\tau(q)$ as a function of $q$ for PRBA, PRBM and Anderson $3d$
models. In models with randomness, to use Eqs. (\ref{pq}),
(\ref{falpha}), and (\ref{falpha2}) we need to take the average of
quantities over (quenched) random sample realizations. To do so we can
take either ensemble average or typical average.  As verified by
Ref. [\onlinecite{ref:vasquez}], typical average yields more accurate
results (since very small numbers in $|\psi_i|^2$ are also take into
account). Thus, we only present the results obtained using typical
averages. To obtain typical average of GIPR for models with disorder,
we rewrite Eq. (\ref{pq}) as:
\begin{equation}
  \e{\langle \ln P_q (\lambda)\rangle} \propto \lambda^{\tau(q)^{typ}},
\end{equation}
where $\langle \cdots \rangle$ stands for arithmetic average over
disorder realization. Thus,
\begin{equation}
  \tau(q)^{typ} = \lim_{\lambda \to 0} \frac{\langle \ln P_q
    (\lambda)\rangle}{\ln \lambda}.
\end{equation}

In taking the limit, we are free to either fix $\ell$ and choose a
sequence of system sizes $N$, or we can fix $N$ and choose a sequence
of smaller values of cell size: $1\ll \ell<N$\cite{ref:mirlin,
  ref:vasquez,ref:rodriguez2008}. Here we choose the former; we choose
$\ell\sim 5$ (for $1d$ models), and $\ell\sim 10$ (for $3d$ model) for
$q<0$ and $\ell=1$ for $q>0$ and we increase the system size $N$. The
reason that we choose $\ell>1$ for negative $q$ is the following:
numerical inaccuracies that are the calculated eigen-mode (either for
$\ket{E_F}$ or MEM) become exaggerated for negative $q$ and thus to
avoid them, we coarse grain over a cell with size $\ell$.  Then, the
slope of the straight line fitting $\langle \ln P_q \rangle$ versus
$-\ln(N)$ gives us the $\tau(q)$.

\begin{eqnarray}
  -\ln(N) \  \tau(q)^{typ} &=& \langle \ln \sum_{k=1}^{N_{\ell}}
                                  \mu_k^q \rangle,
\end{eqnarray}
with similar calculations based on Eqs. (\ref{falpha}) and
(\ref{falpha2}) we obtain $\alpha$ and $f(\alpha)$:
\begin{eqnarray}
  -\ln(N) \ \alpha(q)^{typ} &=&\langle \frac{\sum_{k=1}^{N_{\ell}}
                                    \mu_k^q \ln \mu_k}{P_q}  \rangle\\
  -\ln(N) \ f(q)^{typ} &=&\langle \frac{\sum_{k=1}^{N_{\ell}}
                                    \mu_k^q \ln \mu_k^q}{P_q} - \ln P_q  \rangle
\end{eqnarray}

\begin{figure}
  \centering
  \includegraphics[width=0.48\textwidth]{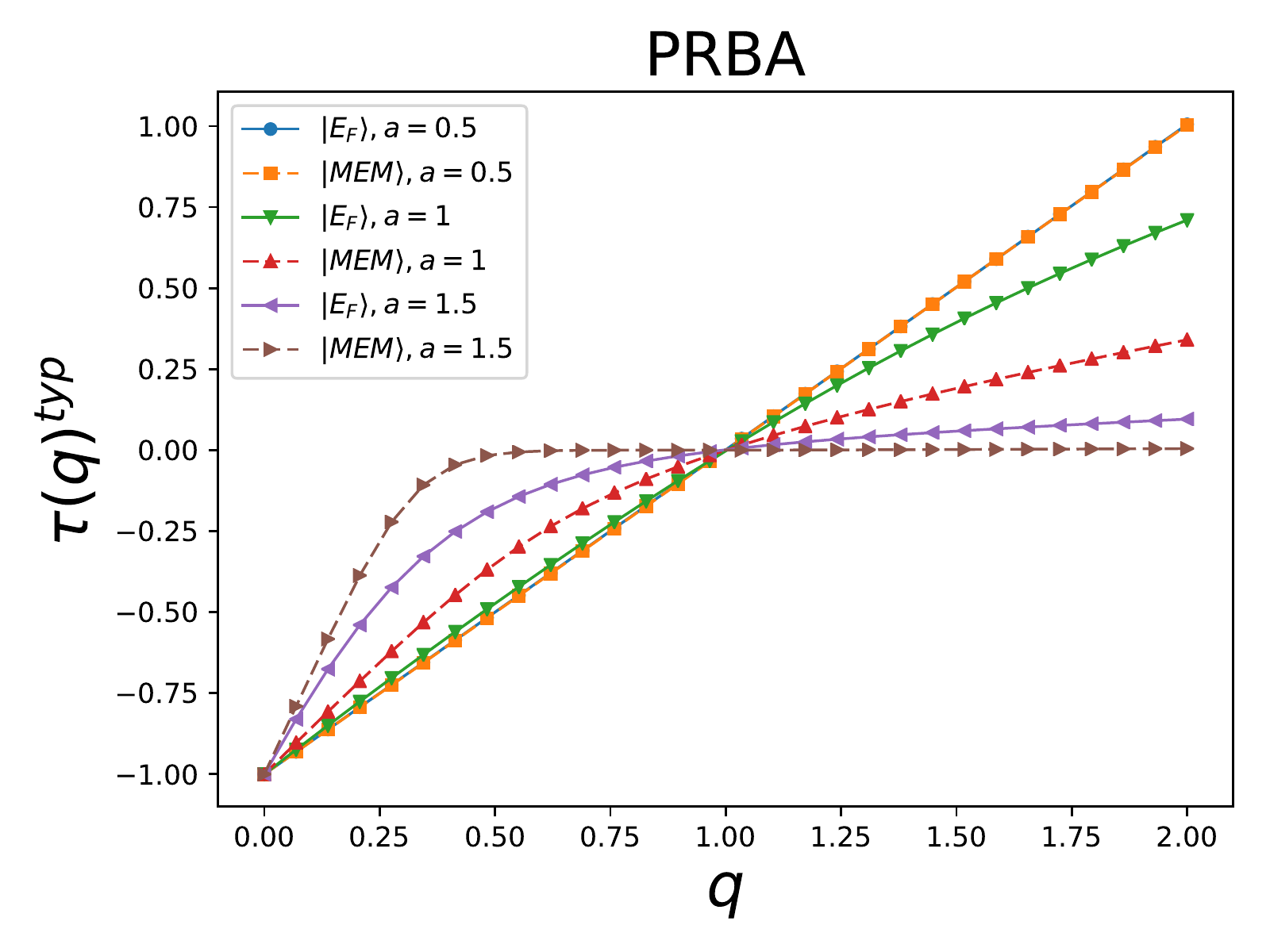}
  \caption{ (color online) Multi-fractality spectrum, $\tau(q)$
    for PRBA. System sizes are between $1000$ and $5000$ in step of
    $500$. For each data point we averaged over $1000$ samples. }
  \label{fig:tau_q_UhNF_MEM_typ_PRBA}
\end{figure}

\begin{figure}
  \centering
  \includegraphics[width=0.48\textwidth]{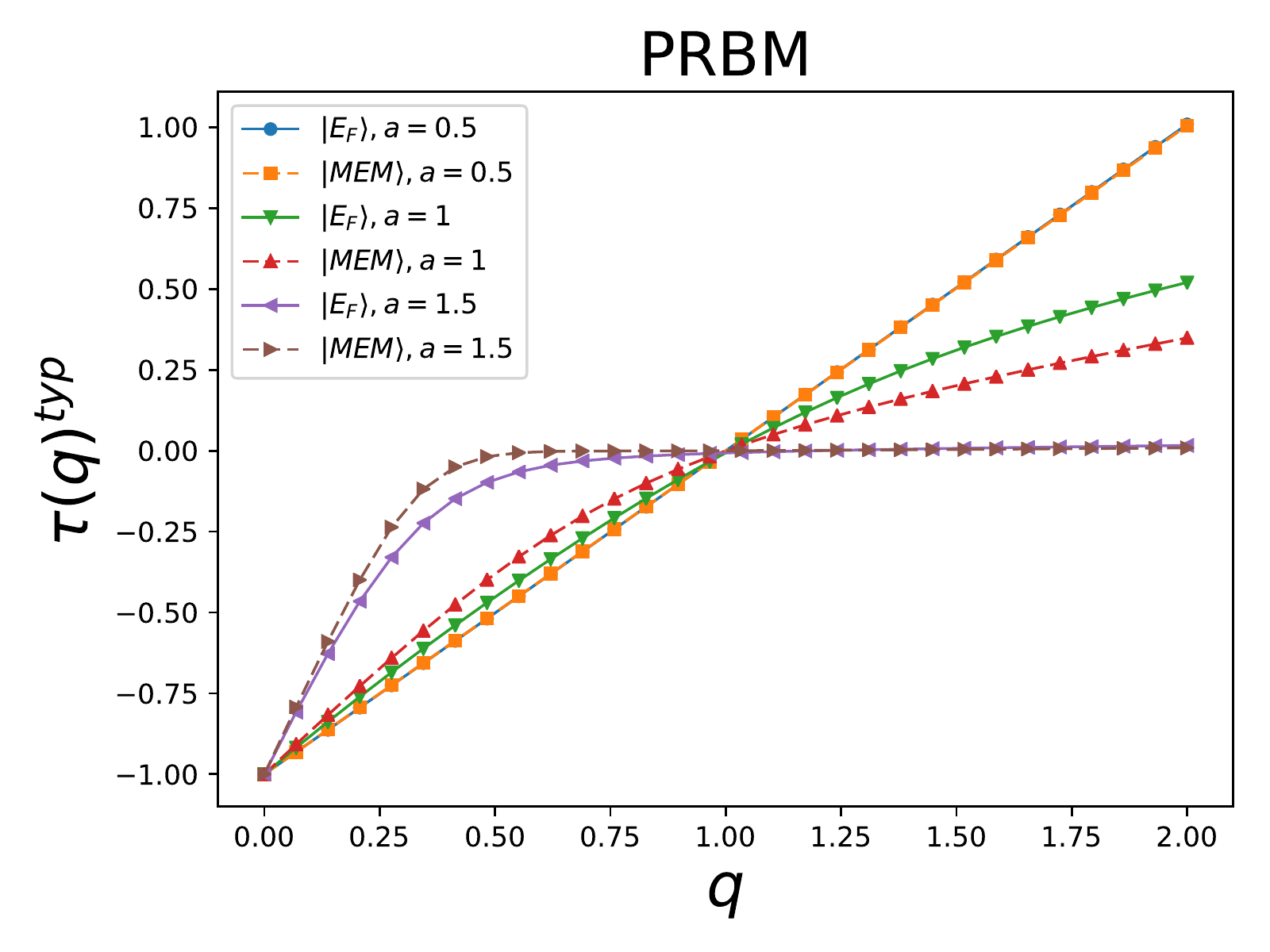}
  \caption{ (color online) Multi-fractality spectrum, $\tau(q)$
    for PRBM. System sizes are between $1000$ and $5000$ in step of
    $500$. For each data point we averaged over $1000$ samples.}
  \label{fig:tau_q_UhNF_MEM_typ_PRBM}
\end{figure}

\begin{figure}
  \centering
  \includegraphics[width=0.48\textwidth]{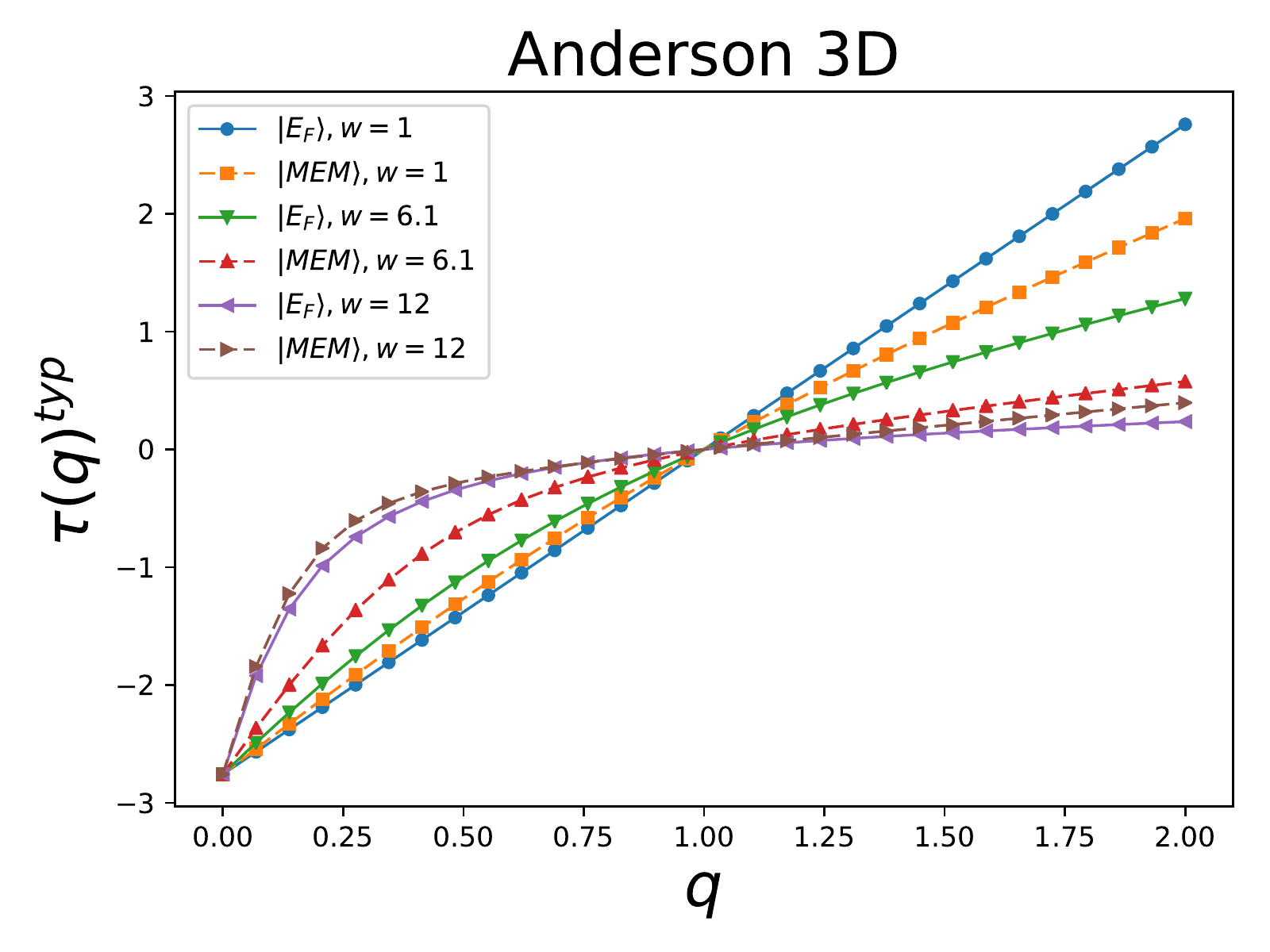}
  \caption{(color online) Multi-fractality spectrum, $\tau(q)$
    for Anderson $3d$ model with Gaussian distribution. System sizes
    are between $4\times 4\times 4$ to $30\times 30\times 30$, with
    $300$ samples for small sizes and $50$ samples for large sizes.}
  \label{fig:tau_q_UhNF_MEM_typ_Anderson3D_G}
\end{figure}

\begin{figure*}
  \centering
  \begin{subfigure}{}%
    \includegraphics[width=0.32\textwidth]{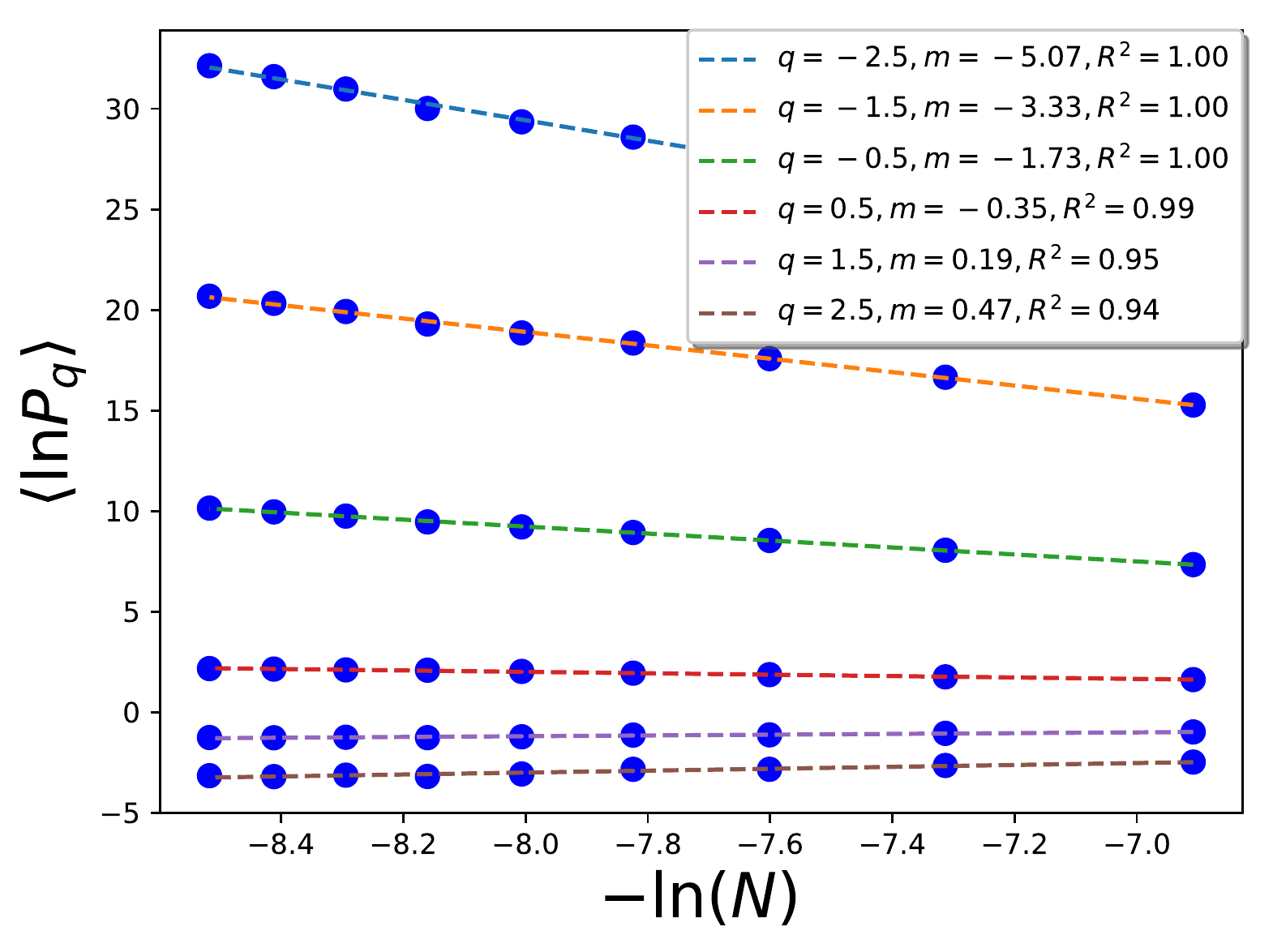}
  \end{subfigure}%
  ~%
  \begin{subfigure}{}%
    \includegraphics[width=0.32\textwidth]{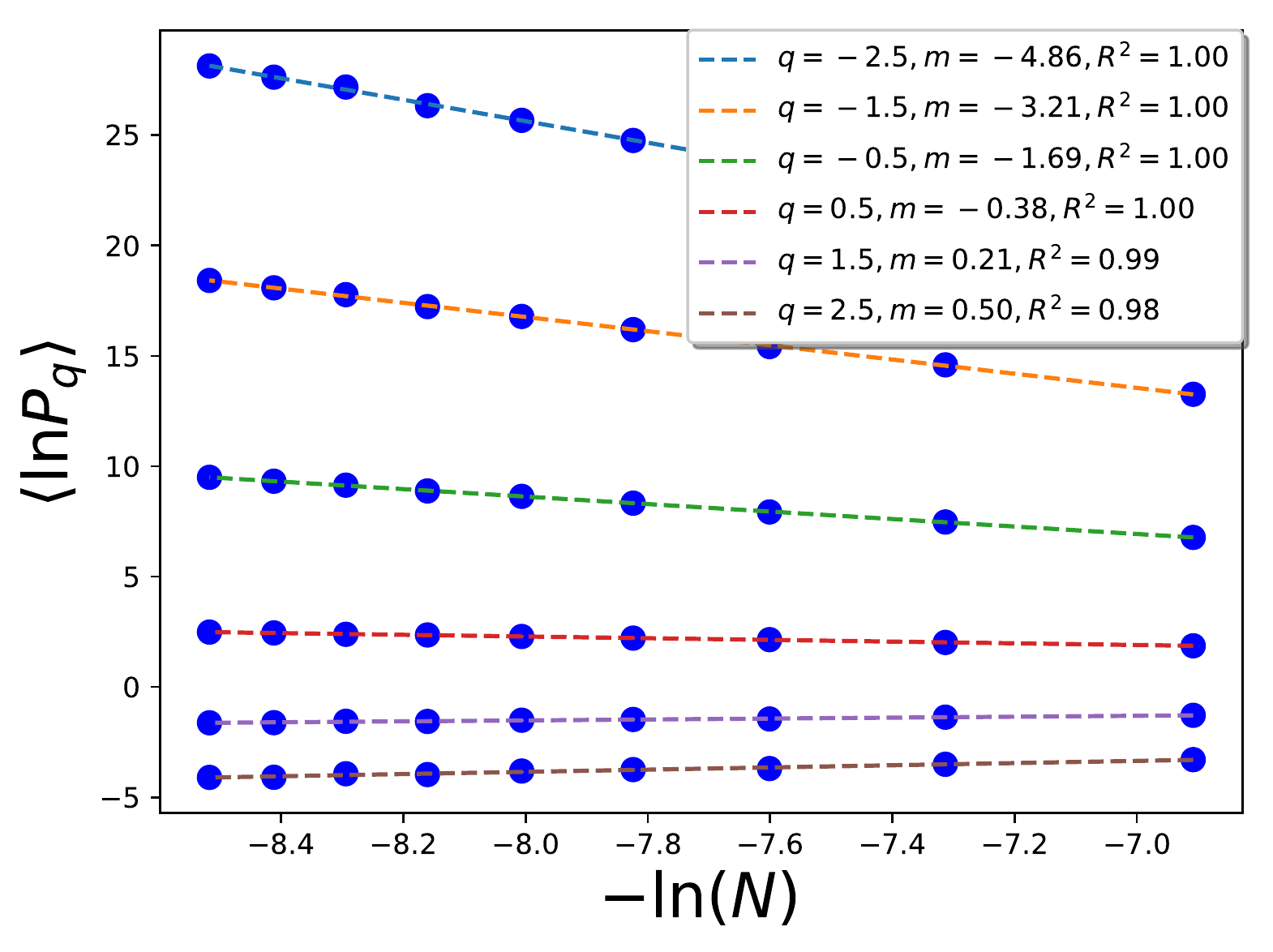}
  \end{subfigure}
  ~%
  \begin{subfigure}{}%
    \includegraphics[width=0.32\textwidth]{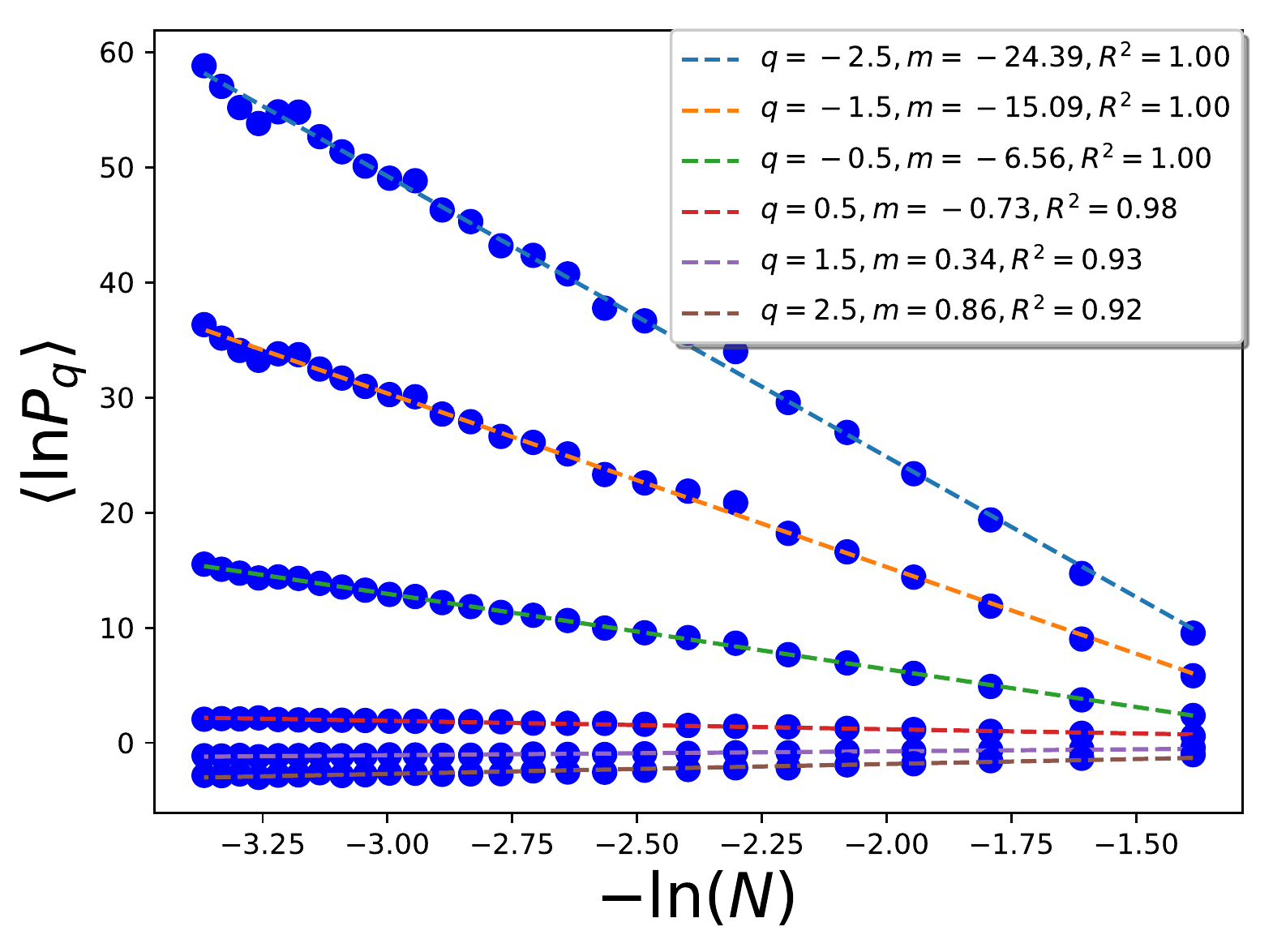}
  \end{subfigure}

  \caption{(color online) Plot of disorder averaged $\ln P_q$ vs $-\ln
    N$ for PRBA(left), PRBM(middle), and Anderson $3d$ (right)
    models for the MEM at the phase transition point. This calculation
    is done for some selected values of $q$ and for each $q$ the slope of the fitted line is indicated by
    $m$. The R-squared which is the sign of how close
    data points are to the fitted line is also calculated (the closer
    to $1$, the better fitted line).
    \label{fig:fitMEM}}
\end{figure*}

We know that, multi-fractal spectrum behavior of Hamiltonian
eigen-mode $\ket{E_F}$ depends on the phase of the system: in the
delocalized phase, $\tau(q)$ is a straight line with a constant slope
equal to dimension of the system. In the localized phase, the spectrum
goes to zero for $q>0$, and at the phase transition point, the slope
of the spectrum is not constant, yielding to multi-fractality. The
multi-fractal spectrum of MEM and $\ket{E_F}$ for PRBA, PRBM, and
Anderson $3d$ model are plotted in
Fig. \ref{fig:tau_q_UhNF_MEM_typ_PRBA},
Fig. \ref{fig:tau_q_UhNF_MEM_typ_PRBM}, and
Fig. \ref{fig:tau_q_UhNF_MEM_typ_Anderson3D_G} respectively. In
Fig. \ref{fig:fitMEM} disorder averaged $\langle \ln P_q \rangle$
versus $-\ln(N)$ plotted and fitted with a straight line for MEM at
the phase transition point for PRBA, PRBM, and Anderson $3d$
models. The slope of this line which is $\tau(q)$ and the accuracy of
the fitted line by R-squared measure are calculated.

In PRBA and PRBM models, the behavior of $\tau$ in the delocalized
phase for both $\ket{E_F}$ and $\ket{MEM}$ are identical, both are
straight lines; although we see a slight discrepancy behavior for
Anderson $3d$ model. In the localized phase, $\tau(q)$ goes to
zero. And, more importantly at the phase transition point, $\tau(q)$
is a non-linear function of $q$.

% \begin{figure*}
%   \centering
%   \begin{subfigure}{}%
%     \includegraphics[width=0.32\textwidth]{PRBA_tau_a1_reg_UhNF.pdf}
%   \end{subfigure}%
%   ~%
%   \begin{subfigure}{}%
%     \includegraphics[width=0.32\textwidth]{PRBM_tau_a1_reg_UhNF.pdf}
%   \end{subfigure}
%   ~%
%   \begin{subfigure}{}%
%     \includegraphics[width=0.32\textwidth]{Anderson_tau_W6p1_reg_UhNF.pdf}
%   \end{subfigure}

%   \caption{(color online) Plot of disorder averaged $\ln P_q$ vs $-\ln
%     N$ for PRBA(left), PRBM(middle), and Anderson $3d$ (right)
%     models for the $\ket{E_F}$ at the phase transition point. This calculation
%     is done for some selected values of $q$ and for each $q$ the slope of the fitted line is indicated by
%     $m$. The R-squared which is the sign of how close
%     data points are to the fitted line is also calculated (the closer
%     to $1$, the better fitted line).
%     \label{fig:fitUhNF}}
% \end{figure*}

\begin{figure*}
  \centering
  \begin{subfigure}{}%
    \includegraphics[width=0.32\textwidth]{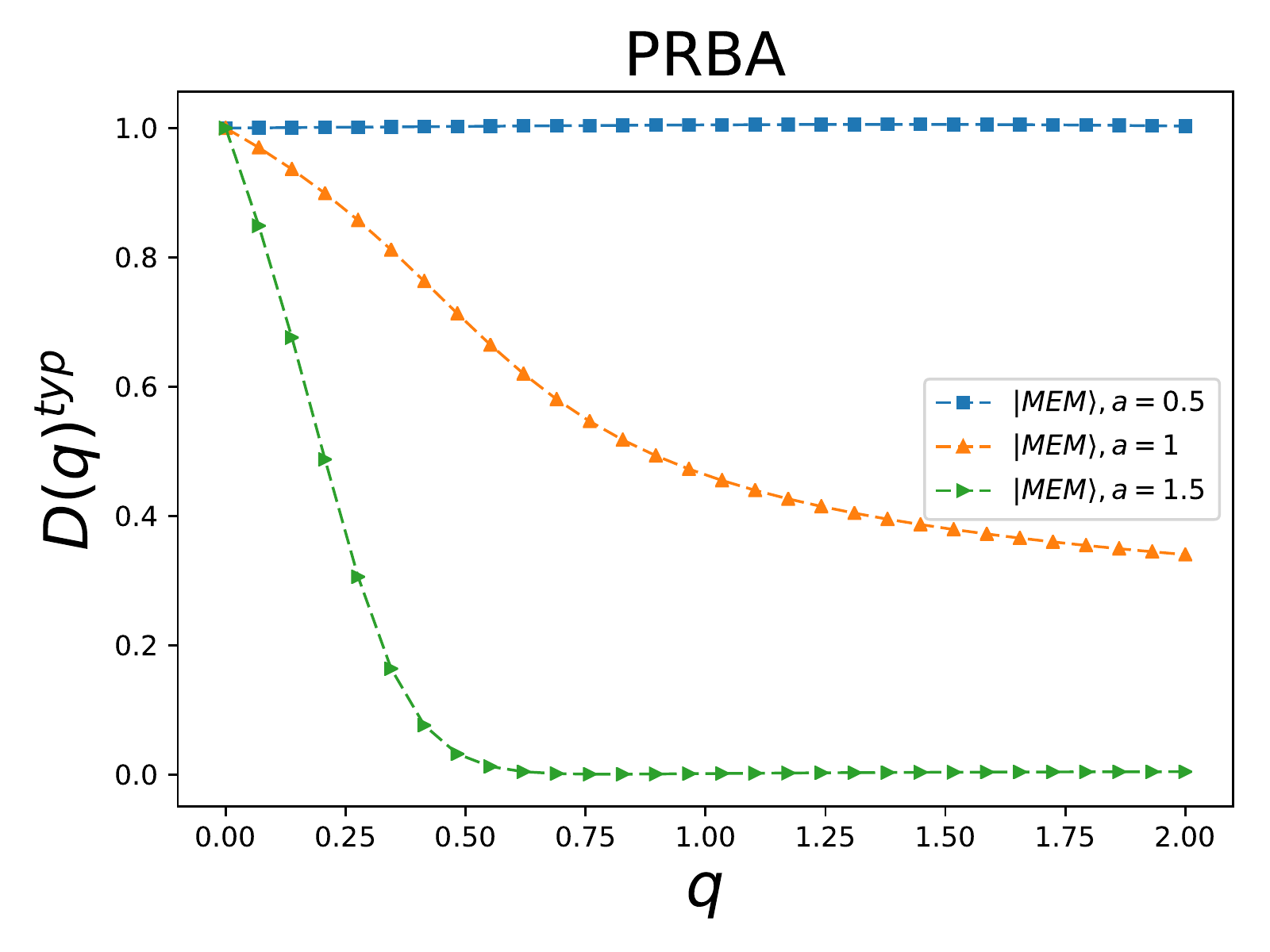}
  \end{subfigure}%
  ~%
  \begin{subfigure}{}%
    \includegraphics[width=0.32\textwidth]{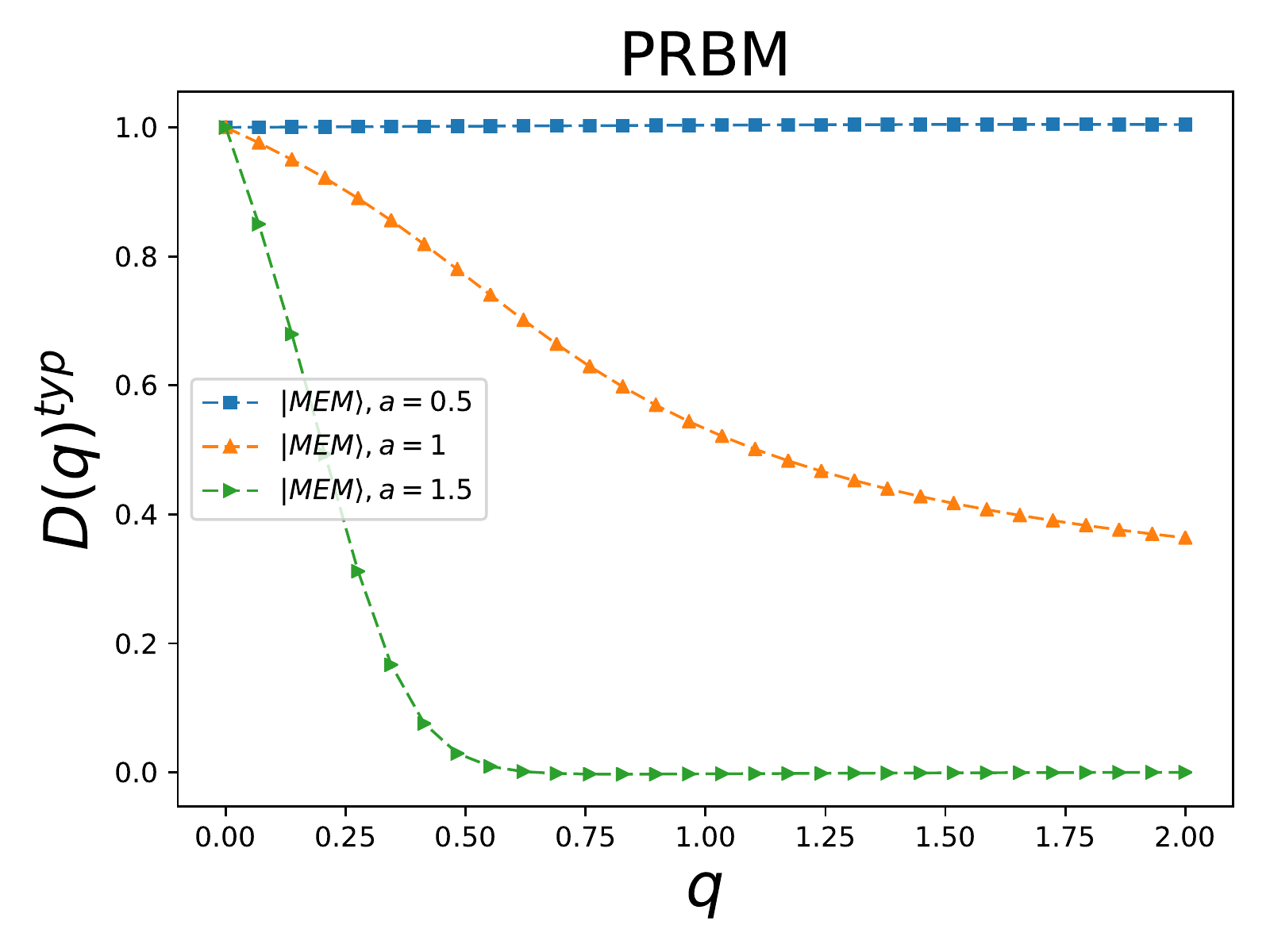}
  \end{subfigure}
  ~%
  \begin{subfigure}{}%
    \includegraphics[width=0.32\textwidth]{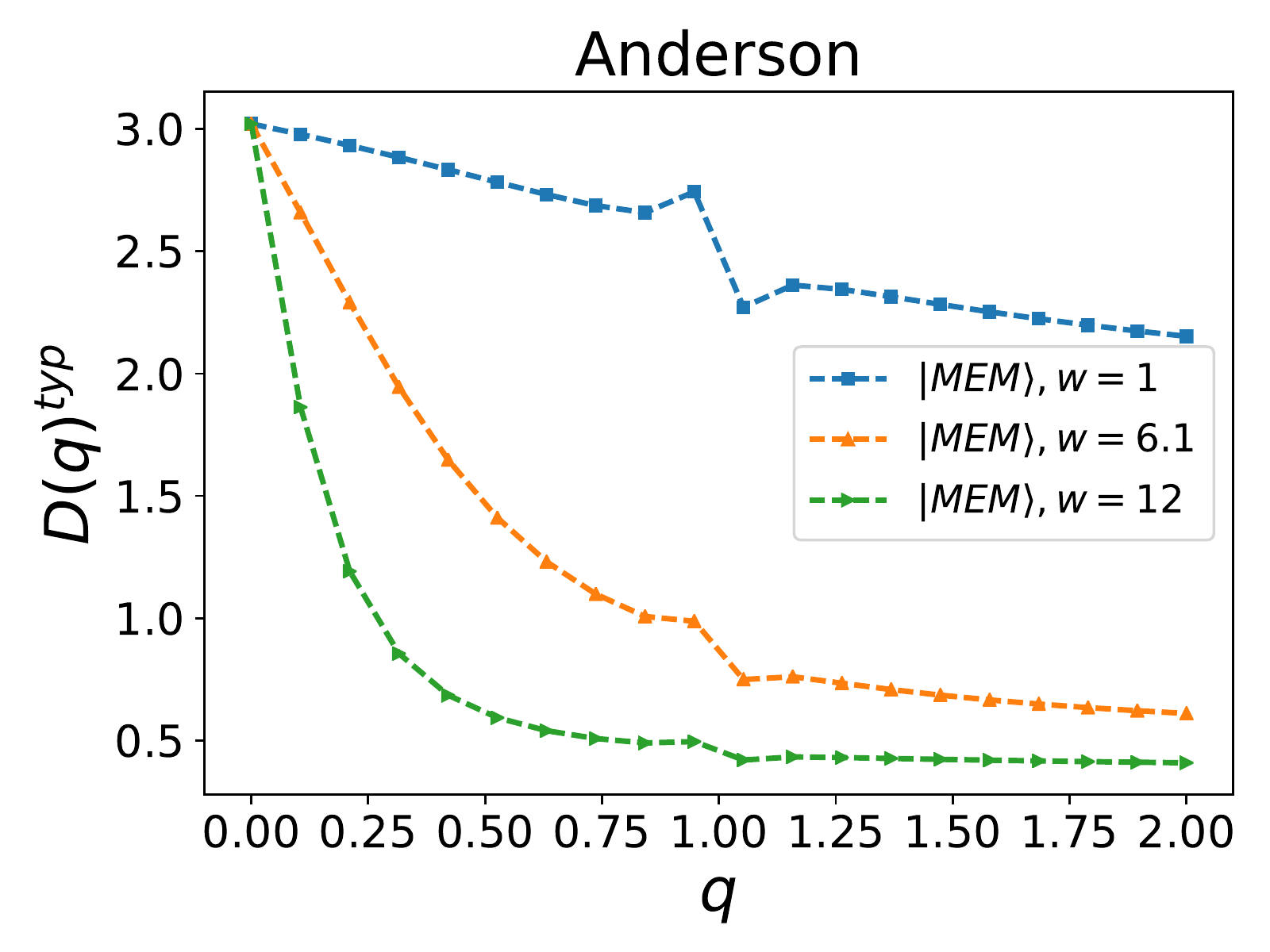}
  \end{subfigure}

  \caption{(color online) Fractal dimension of the MEM of the
    entanglement Hamiltonian for PRBA (left panel), PRBM (middle panel), and
    Anderson $3d$ (right panel) models. For each model, fractal
    dimension is calculated in delocalized phase, at the phase
    transition point, and in the localized phase.
    \label{fig:Dq}}
\end{figure*}

We also calculate the fractal dimension of MEM and plot them in
Fig. \ref{fig:Dq}. The single fractal dimension of MEM in
delocalized phase equals to $1=d$ for PRBA and PRBM models, and
for Anderson $3d$ model, it is around $3=d$.  For the localized
phase, fractal dimension goes to zero as it should. The fractal
dimension of MEM at the phase transition point is also plotted,
which as we can see, is not a constant and thus MEM is
multi-fractal at the phase transition point.

\subsection{Singularity Spectrum}
Next, we consider the behavior of the singularity spectrum $f(\alpha)$
versus $\alpha$. In the delocalized phase, $f(\alpha)$ should be
narrow around $\alpha=d$: when the system is delocalized, we expect
the eigen-mode (either the eigen-mode of the Hamiltonian or the MEM of
entanglement Hamiltonian) to spread over all sites and by the
normalization condition $\sum_{i=1}^{L^d} |\psi_i|^2=1 $, we find that
$ |\psi_i|^2 \sim L^{-d}$. Thus, according to Eq. (\ref{falpha}),
$f(\alpha)$ should be narrowed around $\alpha \sim d$ with the value
of $f(\alpha) \sim d$ (i.e. the fractal dimension of points with
$|\psi_i|^2 \sim L^{-d}$ is very close to the dimension of system
$d$). For PRBA and PRBM $f(\alpha)$ at $\alpha \sim 1$ is close to $1$
and for Anderson $3d$, $f(\alpha)$ at $\alpha \sim 3$ is close to
$3$. At the phase transition point, $f(\alpha)$ has parabolic
behavior\cite{ref:rodriguez2008, ref:mirlin, ref:vasquez} which is the
sign of the multi-fractality of the mode. In the localized phase, the
eigen-mode is localized at a few number of sites and has a very small
value at many other sites, thus $f(\alpha)$ broadens toward larger
$\alpha$, i.e. plot is shifted to the right.

\begin{figure}
  \centering
  \begin{subfigure}{}%
    \includegraphics[trim = 0.2cm  5cm 0 0 clip,width=0.5\textwidth]{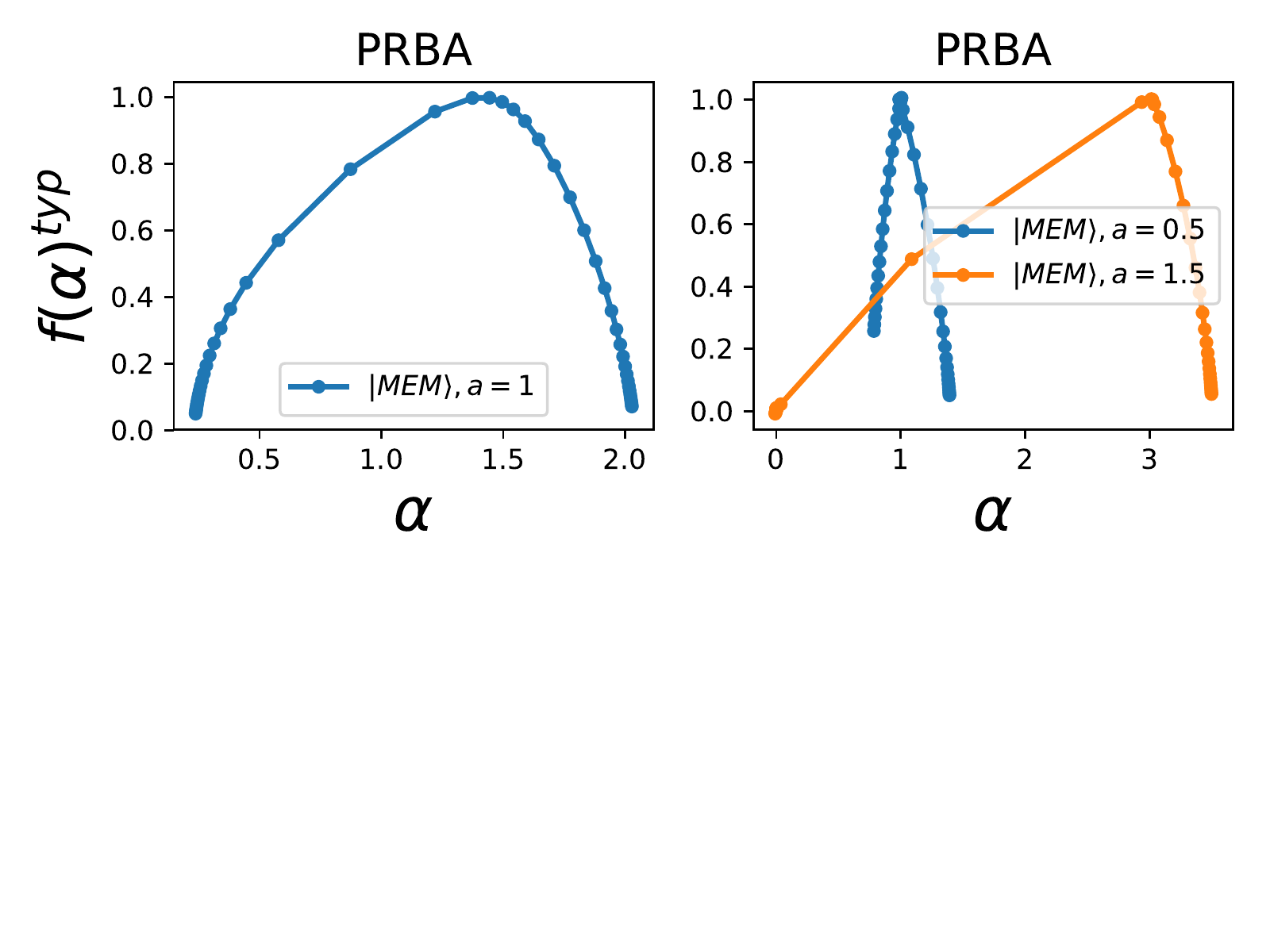}
  \end{subfigure}%
  ~%
  \begin{subfigure}{}%
    \includegraphics[trim = 0.2cm  5cm 0 0 clip,width=0.5\textwidth]{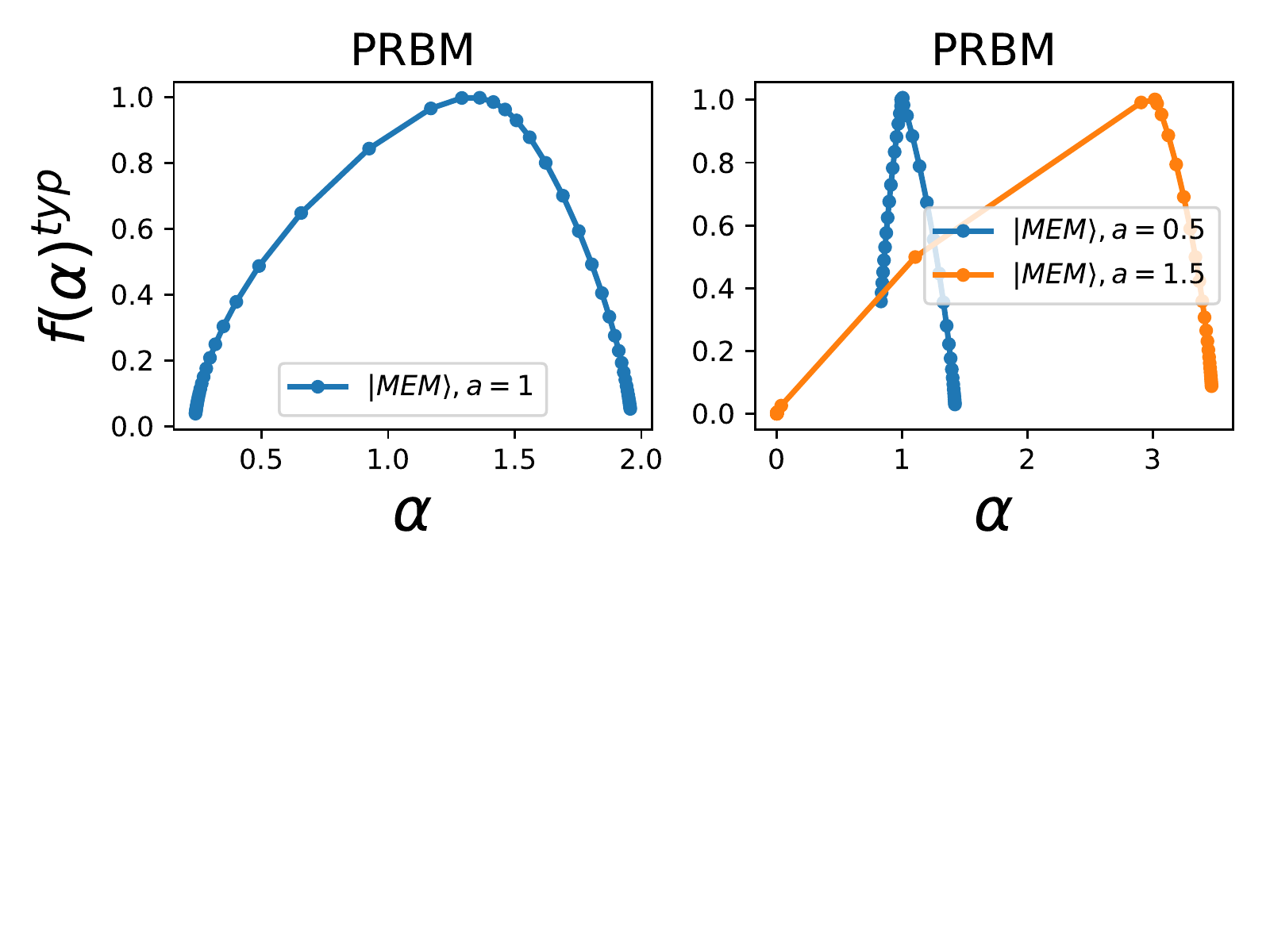}
  \end{subfigure}
  ~%
  \begin{subfigure}{}%
    \includegraphics[trim = 0.2cm  5cm 0 0 clip,width=0.5\textwidth]{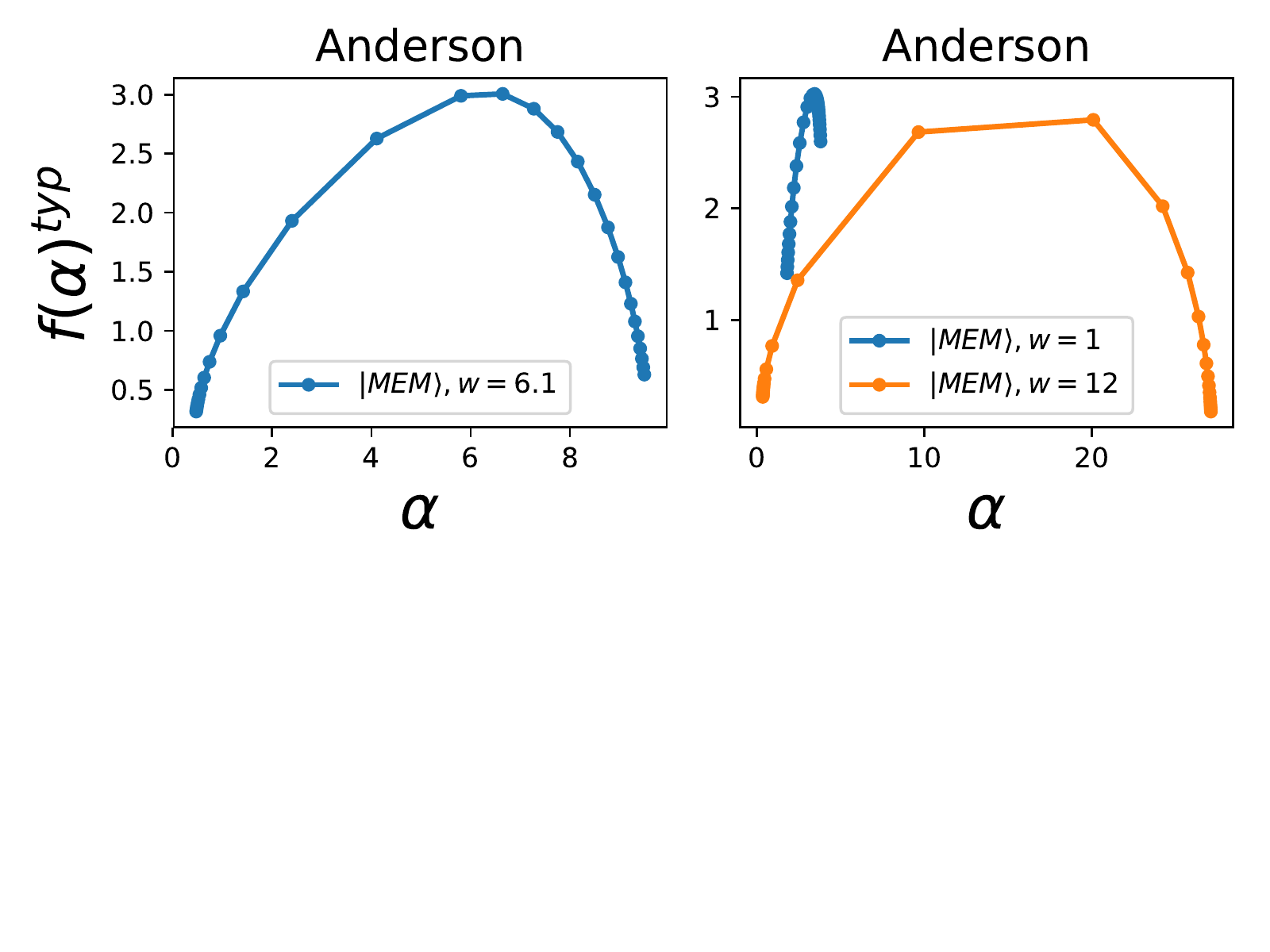}
  \end{subfigure}

  \caption{(color online) Singularity spectrum $f(\alpha)$ for PRBA,
    PRBM, and Anderson $3d$ models. The left panels
    show the singularity spectrum of MEM at the phase transition
    point. The right panels show the singularity spectrum of MEM for
    delocalized and localized phases. For each model, the range of $q$
    is between $-6$ and $6$. 
    \label{fig:falpha}}
\end{figure}

Our calculation of singularity spectrum of MEM for PRBA, PRBM, and
Anderson 3d is plotted in Fig. \ref{fig:falpha}. According to our
calculation, the singularity spectrum of MEM in the delocalized phase
is centered around $\alpha=d=1$ for PRBA and PRBM models and around
$\alpha=d=3$ for Anderson $3d$ model, although it spreads a bit around
$3$ for the Anderson 3d model. At the phase transition point, we see a
parabolic behavior as it is predicted and calculated for Hamiltonian
eigen-mode at the Fermi level. And finally, in the localized phase,
$f(\alpha)$ is broadened toward larger $\alpha$. We note that
$f(\alpha)$ versus $\alpha$ is more broadened in the case of Anderson
$3d$ model than in the $1d$ cases. Beside some inaccuracies that come
from two exact diagonalizations to obtain MEM (as we explained in
Introduction), we expect more broad behavior for the Anderson $3d$
case. Since the linear size of the system, $N$ is the reference in the
calculation of $\alpha$ and $f(\alpha)$ (see Eqs. (\ref{falpha},
\ref{falpha2})), and dimension of the system is three times larger
than the $1d$ cases, thus $\alpha$ goes to larger values. As we can
see, the behavior of the singularity spectrum of MEM like $\ket{E_F}$
depends on the phase considered, and thus it can be used as a phase
detection parameter.

By looking at the Fig. \ref{fig:falpha}, we see that singularity
spectrum at the phase transition point for the three studied models
are symmetric (in contrast to the results obtained in the
Refs. [\onlinecite{ref:mirlin, ref:mirlin2000}] for PRBM and
Ref. [\onlinecite{ref:rodriguez2009}] for Anderson $3d$ models, where
the reason come from choosing $\ell>1$). The physics that is behind
this symmetry(asymmetry) of $f(\alpha)$ versus $\alpha$ which could
indicate uniformities (non-uniformities) in the hierarchical
organization of mode was pointed out in
Ref. [\onlinecite{ref:drozdz}].

\subsection{Symmetry Relation of $\Delta_q$}
\begin{figure*}
	\centering
	\begin{subfigure}{}%
		\def\big{\includegraphics[trim={10pt 0pt 10pt 10pt},clip,width=0.31\textwidth]{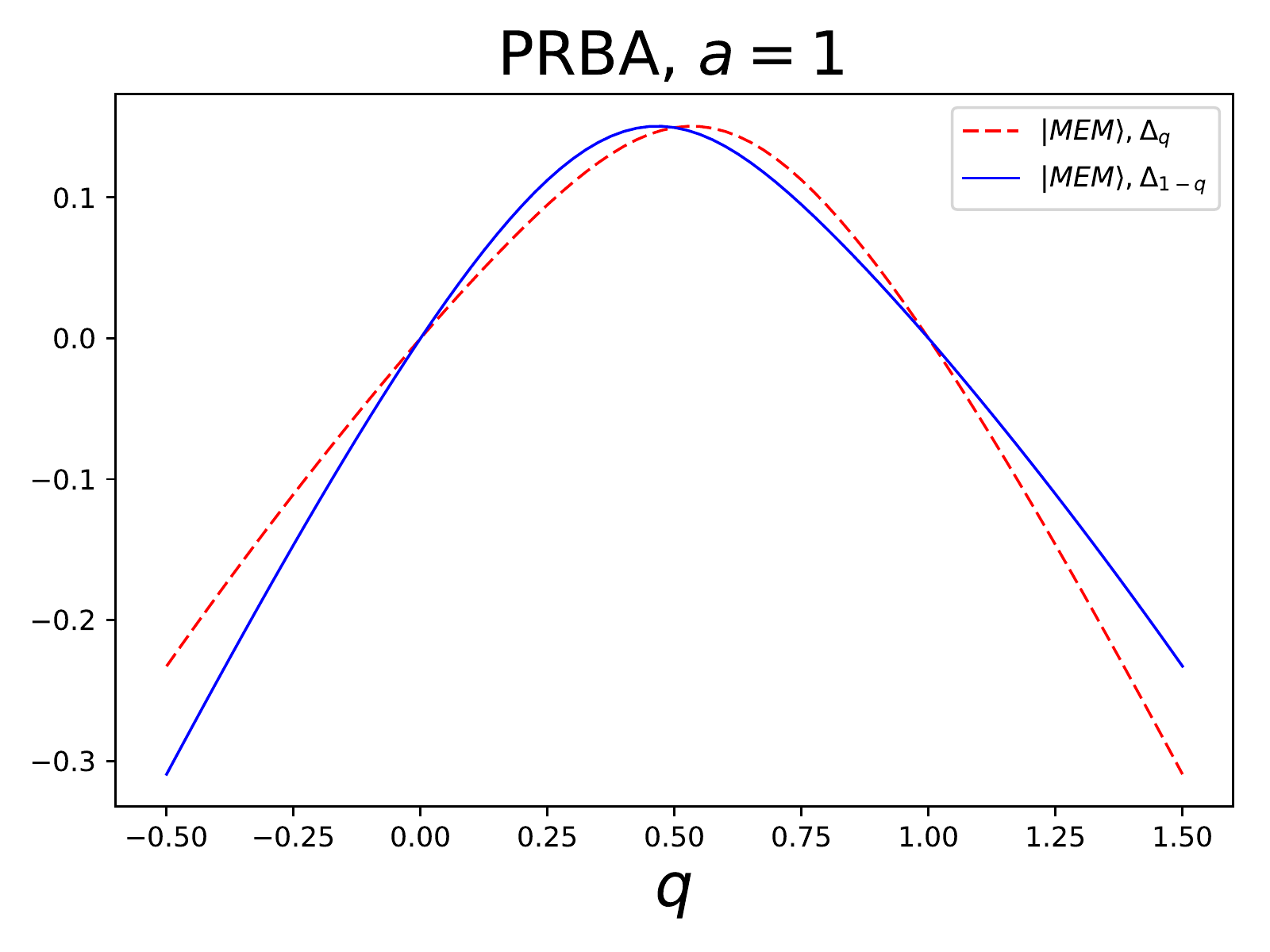}}
		\def\little{\includegraphics[trim={5pt 0pt 10pt 10pt},clip,width=0.14\textwidth]{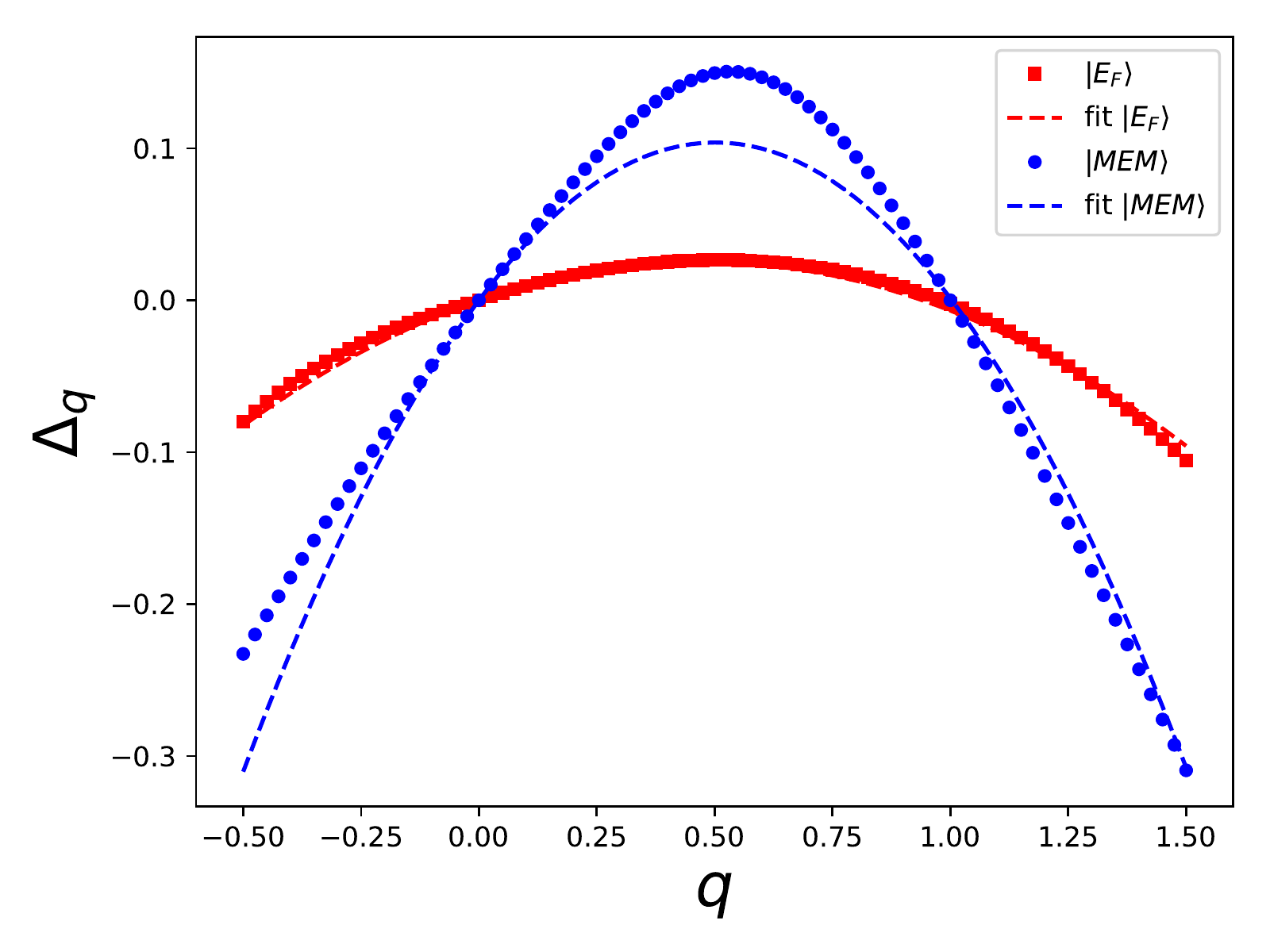}}
		\stackinset{l}{47pt}{b}{20pt}{\little}{\big}
	\end{subfigure}%
	~%
	\begin{subfigure}{}%
		\def\big{\includegraphics[trim={10pt 0pt 10pt 10pt},clip,width=0.31\textwidth]{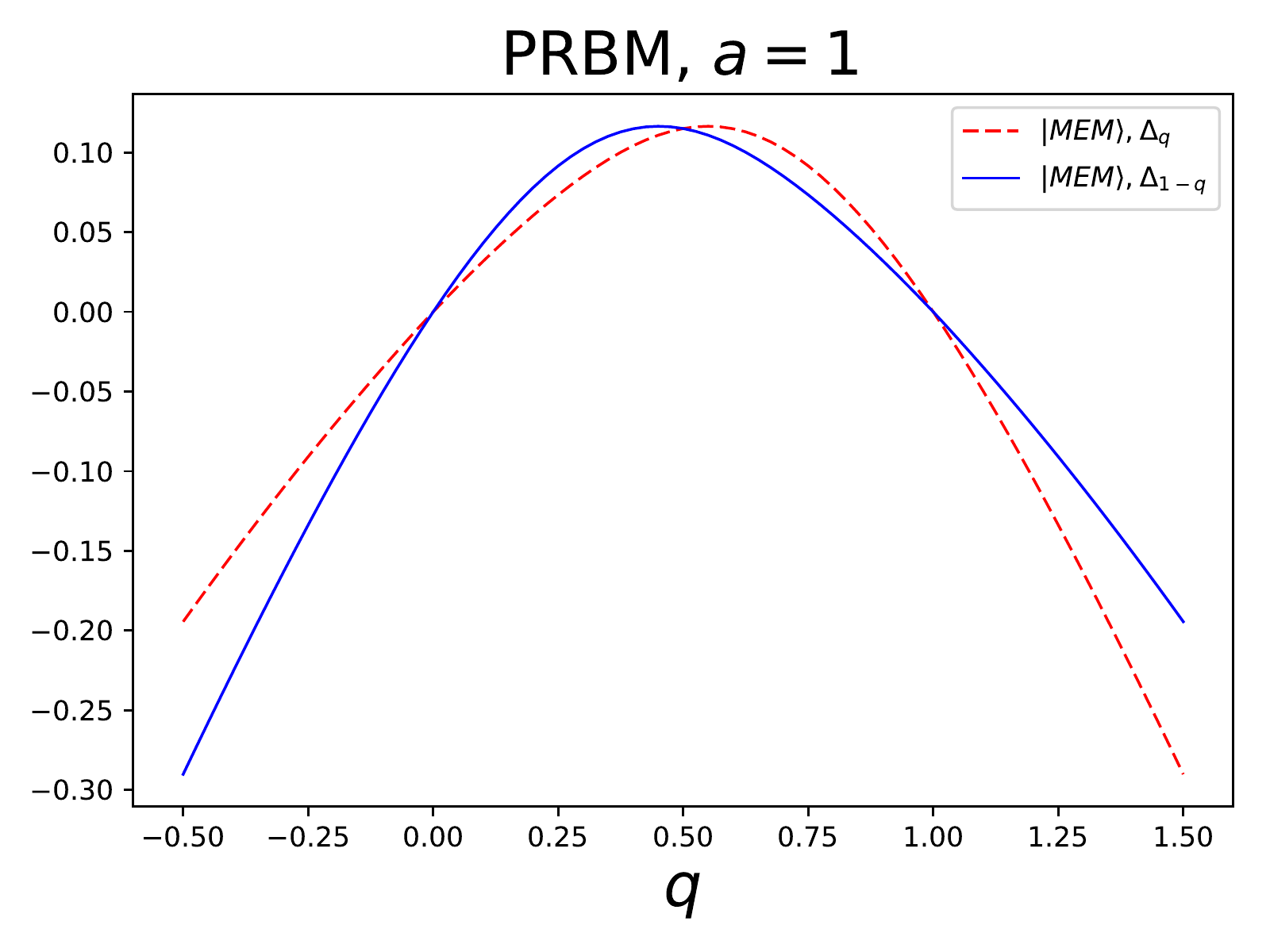}}
		\def\little{\includegraphics[trim={5pt 0pt 10pt 10pt},clip,width=0.14\textwidth]{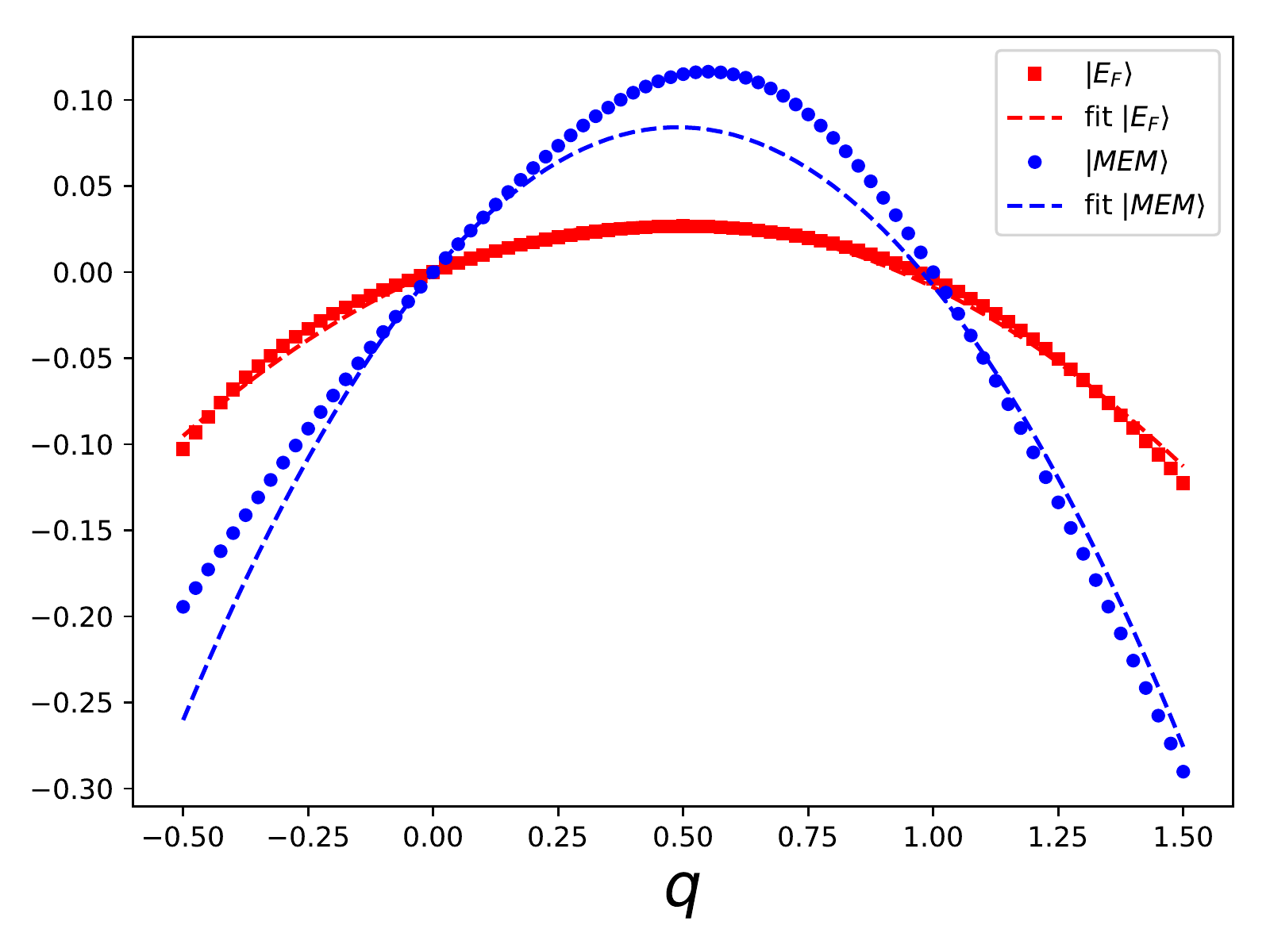}}
		\stackinset{l}{47pt}{b}{20pt}{\little}{\big}
	\end{subfigure}%
    ~%
    \begin{subfigure}{}%
    	\def\big{\includegraphics[trim={10pt 0pt 10pt 10pt},clip,width=0.31\textwidth]{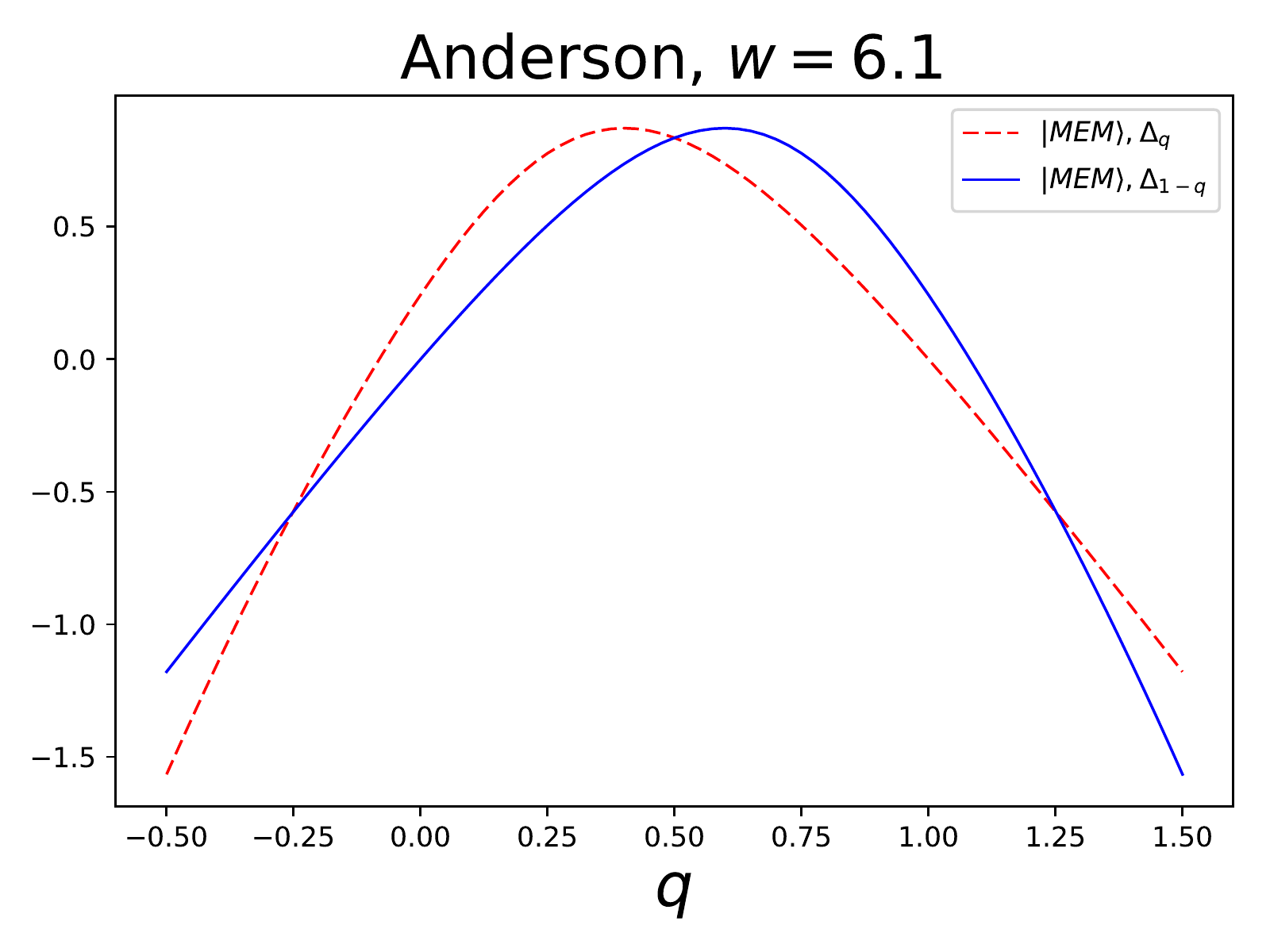}}
    	\def\little{\includegraphics[trim={5pt 0pt 10pt 10pt},clip,width=0.14\textwidth]{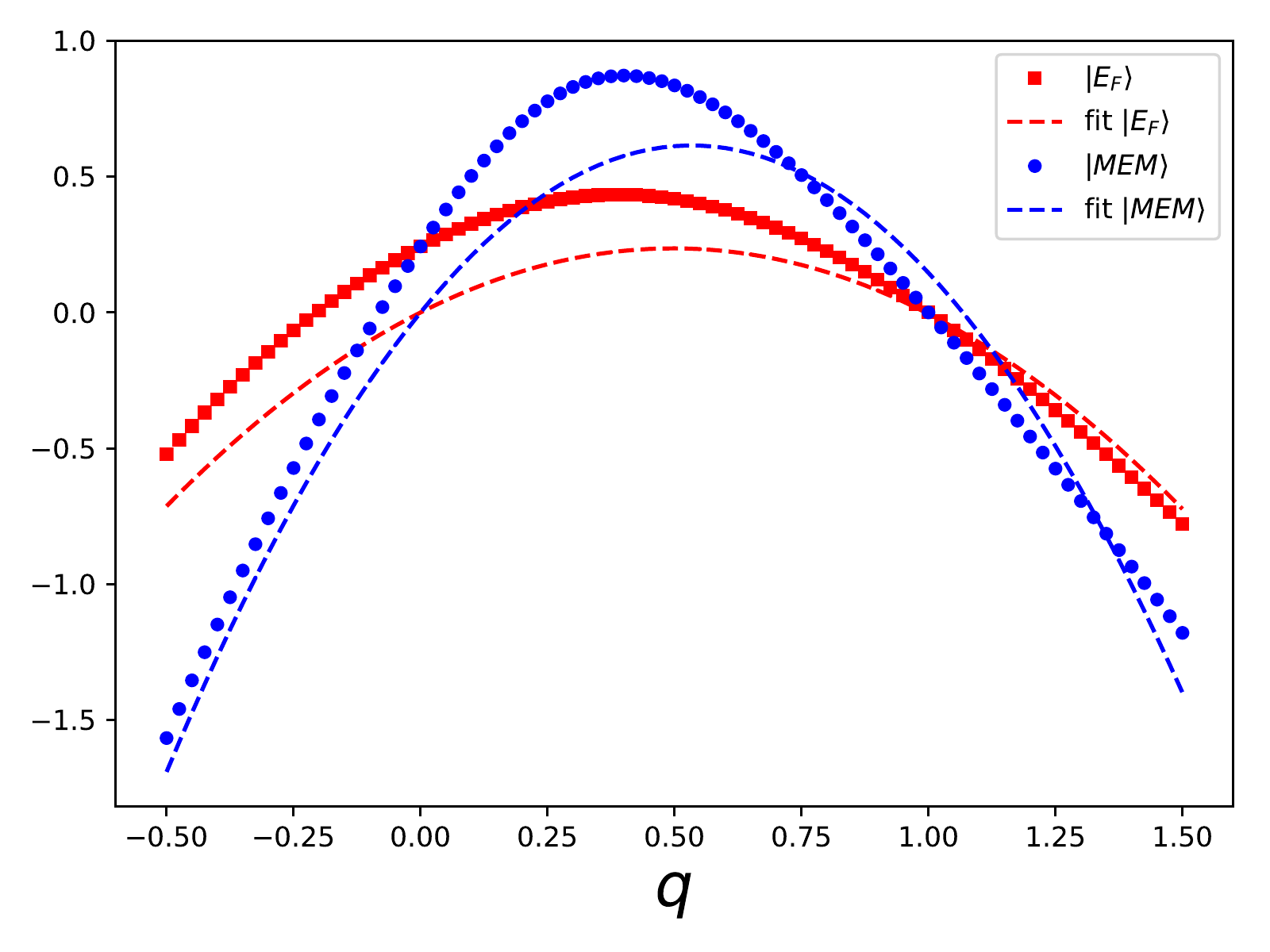}}
    	\stackinset{l}{47pt}{b}{20pt}{\little}{\big}
    \end{subfigure}%
	\caption{(color online) Main panels: both $\Delta_q$ (red)
          and $\Delta_{1-q}$ (blue) are plotted to check the symmetry
          relation $\Delta_q = \Delta_{1-q}$ for MEM in the PRBA, PRBM, and
          Anderson $3d$ models.  Sub-panels: fitting $\Delta_q$ for
          $\ket{E_F}$ (red) and MEM (blue) in each model with
          parabolic equation of $Aq(B-q)$. The values of $A$ and $B$ for $\ket{E_F}$ and MEM  are reported in Table \ref{table:AB}.
          \label{fig:deltaq}}
\end{figure*}

The symmetry relation of anomalous exponents, Eq. (\ref{symDelta}) is
proved analytically and numerically in Refs. [\onlinecite{ref:mirlin,
  ref:vasquez}] for $\ket{E_F}$. Here we present numerical
verification of the symmetry relation for the MEM in PRBA, PRBM, and
Anderson $3d$ models in the main panels of Fig. \ref{fig:deltaq}. As
we can see the symmetry relation of Eq. (\ref{symDelta}) is respected
for the MEM of the entanglement Hamiltonian.  On the other hand, we
fit the $\Delta_q$ for the $\ket{E_F}$ and MEM with the parabolic
equation of $Aq(B-q)$ and find the $A$ and $B$ constants. The values
of $A$ and $B$ are reported in the Table \ref{table:AB}. For three
models considered, and for both $\ket{E_F}$ and MEM, $B \sim 1$ as it
should be (since $\tau(1) = 0 \to \Delta_1 = 0$, and so
$B=1$). Moreover, $A_{\text{MEM}}$ is approximately three times larger
than $A_{\ket{E_F}}$ in each model.

\begin{table}[H]
  \caption{\label{table:AB} Constants $A$ and $B$ when we fit
    $\Delta_q$ for $\ket{E_F}$ and MEM with the equation $Aq(B-q)$
    (see sub-panels of Fig. \ref{fig:deltaq}). }
	\centering
	\begin{tabular}{c|ccc}
		    & PRBA $\ \  $   &  PRBM $\ \  $ &  Anderson $3d$ \\   
		\hline
		$A_{\ket{E_F}}$ & 0.11(4) & 0.13(3) & 0.9(5)\\
		
		$A_{\text{MEM}}$ &  0.4(1) &  0.35(2) &  2.1(6) \\
		
		$B_{\ket{E_F}}$ & 0.94(5) & 0.93(3) & 0.9(9)\\
		
		$B_{\text{MEM}}$ & 1.0(0) & 0.97(8) & 1.0(7)
	\end{tabular}
\end{table}

\section{Concluding Remarks}\label{conc}

It has been shown that Anderson transition as a quantum phase
transition exhibits multi-fractal behavior at the critical point.  In
fact, the generalized participation ratio of Hamiltonian eigen-mode at
the Fermi level is a measure that shows multi-fractality of the
system. Recently, entanglement Hamiltonian and its associated
maximally entangled mode has attracted attention as a tool to
characterize systems behavior particularly at the critical point. We
note that obtaining an explicit relation for eigenvectors of the
entanglement Hamiltonian (EH) based on the eigenvectors of Hamiltonian
is not trivial and they are not directly related. In a
study\cite{ref:eisler}, people found the explicit expression for the
EH matrix elements in the ground state of free fermion models. People
also found that at the extreme limit of strong coupling between two
chosen subsystems, EH of a subsystem and its Hamiltonian are
proportional\cite{ref:peschel}. In this paper, we have shown that
multi-fractality of Anderson transition carries over to MEM much in
the same way as Hamiltonian eigen-mode at the Fermi level.

Based on numerical calculations for PRBA and PRBM $1d$ models, and
also Anderson $3d$ model, we showed that single particle MEM of the
entanglement Hamiltonian, has the same fractal properties as the
Hamiltonian eigen-mode at the Fermi level; although for Anderson $3d$
model we see a little deviations, since we could not reach very large
system sizes. For MEM, in the delocalized phase, $\tau(q)$ has a slope
equal to the dimension of the system, while in the localized phase, it
goes to zero. Interestingly, at the phase transition point, MEM is
multi-fractal and its multi-fractality is similar to that of the
$\ket{E_F}$. MEM also follows the symmetry relation of anomalous
exponents. Moreover, singularity spectrum $f(\alpha)$ of MEM, is
similar to $f(\alpha)$ of $\ket{E_F}$: in the delocalized phase it is
around $\alpha \sim d$; at the phase transition point it has parabolic
shape with the maximum value $d$, and it broadens in the localized
phase. And thus, by looking at multi-fractal spectrum or singularity
spectrum of MEM, we can distinguish different phases.

Multi-fractality of an observable at the quantum phase transition
means that this observable is self-similar; and finite-size scaling of
observable is a legitimate method of obtaining critical
exponents. Here we saw that entanglement Hamiltonian shows
multi-fractality; which indirectly verifies that reduced density
matrix and even entanglement entropy should exhibit self-similarity at
the phase transition point, and thus their finite size scaling can be
used as a method to calculate the critical exponents, as it has been
done in Refs. [\onlinecite{ref:zhao, ref:igloi, ref:wang}].

Multi-fractality of entanglement Hamiltonian was studied in this paper
through its eigen-modes. This study could also be done by inspecting
the multi-fractality of elements of entanglement Hamiltonian. It might
also be interesting to consider entanglement entropy, or a measure of
eigenvalues of reduced density matrix, to see if they also carry
signatures of multi-fractality.

\section{Acknowledgments}
This research is supported by University of Mazandaran, National Merit
Foundation of Iran, and Institute for Research in Fundamental Sciences
(IPM). Part of this work was done while I was working at Shiraz
University. I would like to thank Dr. Abbas Ali Saberi and Afshin
Montakhab for useful discussions and their constructive criticism on
the manuscript.

\end{document}